\documentclass[twocolumn,showpacs,
showkeys,superscriptaddress,prc,nofootinbib]{revtex4-2}
\usepackage{hyperref}
\usepackage[hyphenbreaks]{breakurl}
\hypersetup{breaklinks=true,colorlinks=true, citecolor=blue, urlcolor=blue, linkcolor=blue}
\PassOptionsToPackage{hyphens}{url}
\usepackage{CJK}
\usepackage{mathrsfs}
\usepackage{amssymb}
\usepackage{amsmath}
\usepackage{graphicx}
\usepackage{dcolumn}
\usepackage{bm}
\usepackage{graphicx}
\usepackage{float}
\usepackage[normalem]{ulem}
\usepackage{color}
\usepackage{multirow}
\usepackage{enumerate}
\usepackage{ulem}
\usepackage{textcomp}

\newcolumntype{d}[1]{Dc{.}{.}{#1}}

\begin{document}
	\begin{CJK*}{UTF8}{}
		
		\title{
			Robust correlation between binding energies and charge radii of mirror nuclei
		}
		\author{Y. Lei ({\CJKfamily{gbsn}雷杨})}
		\email[]{leiyang19850228@gmail.com}
		\affiliation{School of Nuclear Science and Technology, Southwest University of Science and Technology, Mianyang 621010, China}
		\author{N. A. Alam}
		\affiliation{School of Mathematics and Physics, Southwest University of Science and Technology, Mianyang 621010, China}
		\author{Z. Z. Qin ({\CJKfamily{gbsn}覃珍珍})}
		\affiliation{School of Mathematics and Physics, Southwest University of Science and Technology, Mianyang 621010, China}
		\author{M. Bao ({\CJKfamily{gbsn}鲍曼})}
		\affiliation{Department of Physics, University of Shanghai for Science and Technology, Shanghai 200093, China}
		\author{K. Y. Zhang ({\CJKfamily{gbsn}张开元})}
		\affiliation{Institute of Nuclear Physics and Chemistry, China Academy of Engineering Physics, Mianyang 621900, China}
		\author{C. Ma ({\CJKfamily{gbsn}马畅})}
		\affiliation{Shanghai Key Laboratory of Particle Physics and Cosmology, School of Physics and Astronomy, Shanghai Jiao Tong University, Shanghai 200240, China}
		\date{\today}
		
		\begin{abstract}
			Using the charge density from the two-parameter Fermi model, 
			{\color{blue}
			a robust and nontrival correlation between binding energis and charge radii of mirror nuclei is newly proposed. This correlation enables simple yet reliable predictions of the nuclear mass and charge radius of  proton-rich nuclei.
			}
		 	The validity of these predictions is demonstrated by comparing the predicted binding energies and charge radii with experimental data and predictions from other models. All 197 predicted binding energies and 199 charge radii involved in the comparisons are tabulated in the Supplemental Materials of this paper. The noticeable discrepancies are attributed to the large asymmetry in charge densities of mirror nuclei, suggesting that the proposed correlation could be a sensitive probe for local structural anomaly, such as shell closure and proton halo. 
		 	{\color{blue}
		 	The difference in mass dependence of charge radii near the proton dripline compared to those along the $\beta$-stability line supports the validity of our prediction method.
		 	}  
		\end{abstract}
		\maketitle
	\end{CJK*}
	
	\section{Introduction}\label{sec-int}
	As the new generation of radioactive ion beam facilities begins operation and construction \cite{meng1,meng2,meng3,meng4,meng5,meng6}, more and more nuclei far from stability are being and will continue to be produced. This extends the nuclear landscape from stable to exotic nuclei \cite{meng7} and provides new insights into nucleosynthesis in stars and stellar explosions \cite{meng8,meng9}.
	Nuclear mass (or mass related observable like nucleon separation energy, $\alpha$ decay energy) is usually the first observable measured for a newly discovered nuclide. Moreover, masses of proton-rich nuclei play vital roles in the rapid proton capture process and neutrino-induced nucleosynthesis \cite{wang5,wang6}. Therefore, accurate mass predictions for proton-rich nuclei are highly desirable for studying proton-rich nuclide synthesis in the lab and in stars.
	
	Mass predictions are typically based on global nuclear structure models or local mass correlations \cite{LD2003}. Global models have been successfully developed since the 1930s, beginning with Gamow's proposal of the liquid drop model for nuclear binding energy \cite{bao16}. Building on the liquid drop model, the Bethe-Weizs\"acker formula provides a simple yet profound method for estimating nuclear binding energies \cite{bao17,bethe}. Beyond the simple macroscopic liquid drop model, nuclear microscopic structure is introduced into macroscopic-microscopic models to further improve nuclear mass descriptions, such as the finite range droplet model \cite{bao19,bao20,uncertainty-method,bao22}, the Weizs\"acker-Skyrme model \cite{bao23,bao24,bao25,bao26,bao27}, and their variants \cite{bao28}. Microscopic structural models are also used for mass analysis and prediction, such as the Duflo-Zuker model \cite{wang7}, relativistic mean-field theory \cite{RMF1999}, the Skyrme Hartree-Fock-Bogoliubov model (e.g., Refs. \cite{bao31,bao34}), relativistic continuum Hartree-Bogoliubov (RCHB) theory \cite{bao35,bao7}, and deformed relativistic Hartree-Bogoliubov theory in continuum (DRHBc) \cite{bao36,DRHBc-model}.

	Recently, the DRHBc Mass Table Collaboration presented the ground-state properties of even-$Z$ nuclei across the nuclear chart, successfully reproducing nuclear masses within an RMS (root mean square) deviation of 1.433 MeV and charge radii within approximately 0.03 fm \cite{DRHBc22,DRHBc24}. However, global models are optimized for the entire nuclide chart, where drip-line nuclei have little influence. As a result, global models are considered less accurate for drip-line nuclei. In contrast, some mass correlations, such as Audi-Wapstra systematics (see \cite{AME202} for example), the Gareyy-Kelson relations \cite{bao44,bao45}, and the systematics of residual proton-neutron interaction \cite{bao46,bao47,bao48,bao49}, have been successful in making precise mass predictions at the edge of known nuclei but usually fail to predict the masses of drip-line nuclei for two reasons. First, these correlations are typically valid for several nearby nuclei. As the distance between nuclei increases, their correlation weakens, introducing significant systematic errors. However, predicting the masses of drip-line nuclei requires a long-range correlation that usually extends from the valley of stability to the drip line. Second, if one insist on a short-range correlation for drip line mass prediction, such a correlation must be applied iteratively, causing prediction uncertainty to accumulate rapidly. Therefore, it is desirable to establish a long-range correlation involving fewer nuclei.
	
	The isospin symmetry enables such a long-range correlation. Isospin symmetry arises from the charge independence of the nuclear force, which dictates that the binding energy difference between mirror nuclei is primarily due to the Coulomb force. Compared to the nuclear force, the Coulomb force is better understood and can be easily modeled with a simple but reasonable charge density. This allows the binding energy difference between mirror nuclei to be calculated without the detail of nuclear structure governed by the nuclear force, making it possible to predict the binding energy of a proton-rich nucleus using just one experimental input: the known binding energy of its mirror nucleus. More importantly, the mirror mass relation is independent of distance, enabling the establishment of a long-range correlation between mirror nuclei with larger proton-neutron number differences. This approach has recently been adopted to predict the binding energies of proton-rich nuclei such as in the isobaric multiplet mass equation \cite{WS1959,G1969,BW1979,JB1998}, improvements to the Garvey-Kelson mass relation \cite{GK1966,zong16,zong17}, and mass relations of mirror nuclei \cite{zong18,zong19,zong20,zong21,zong-self}. Most recently, additional corrections have been made to the mass relations of mirror nuclei, taking into account the global impact of pairing and shell effects \cite{FU2024,BAO2024}.
	
	The nuclear charge radius is another key observable for studying nuclear structure. For example, it is related to shell closure and weakening \cite{rc13,wang-l,rc213,Sc-self,Sc-6,Sc-8,Sc-9,Sc-12,Sc-15,Sc-20,Ca-1-12,K-self}, 
	{\color{blue}
	pairing \cite{Sc-8,K-9-coex-pair,K-10-pair}, nuclear shape \cite{Sun-self,Sun-shape-11,K-7-shape-staggering,Sun-shape-with-zhao} and shape coexistence \cite{K-8-shape-coex,K-9-coex-pair}, the nuclear equation of state \cite{wang-l},
	}
	 exotic neutron distributions and halo structures \cite{rc214,K-6-halo}, and isomer shift \cite{rc216}. The recent development of new radioactive ion beam facilities has enabled measurements of the charge radii of unstable nuclei via isotope shifts from laser spectroscopy experiments (for example, see Refs. \cite{rc219,rc2110,Sc-20,Sc-8,Sc-6,isotope-shift,Sc-self,Sc-8,Ca-1-18,Ca-1-19,K-self}), 
	 {\color{blue} Additionally, a new compilation of experimental charge radii has been developed \cite{rc21}. These advances greatly expand the range of charge radius data, and makes access to new data easier.} 
	 	
	 Similarly to nuclear mass, there also exist global and local models for charge radii (for example, see Refs. \cite{bao29,bao30,bao32,bao33,bao68,bao69,Z2010,B2008} and \cite{bao71,Sun-shape-with-zhao,bao73,Sun-self,bao75,bao76,Li2023}, respectively), which certainly advance our understanding of nuclear charge structure. In particular, the difference in mirror pair charge radii has been suggested as a probe for the slope of the symmetry energy (for example, see Refs. \cite{wang-l,brown-l,yang-l,npa-l,brown-l-2,prl-l}).

	Since both the mass difference of mirror nuclei and their charge radii are related to the nuclear electric system, it is possible to establish a correlation between the mass and radius of mirror nuclei, by assuming a reasonable charge density. {\color{blue} Historically, a correlation of this kind could have been derived from Bethe's proposal for estimating nuclear size \cite{janecke22,rmp}. Moreover, any sophisticated nuclear structure model could numerically connect the properties of nuclear densities and binding energies, as demonstrated in Ref.  \cite{li-data-driven}. However, we have not yet encountered an explicit correlation for mirror nuclei in the existing literature. Thus, it is new to explicitly establish an analytical correlation between charge radii and binding energies for mirror nuclei.}
		
	In this work, we combine nuclear binding energies and charge radii of mirror nuclei into a single correlation. A new observable in this correlation can be quantitatively estimated, based on certain nuclear charge density. We demonstrate that this observable is insensitive to the details of charge density and varies little across different regions. Therefore, the resulting correlation is robust, which may help predict unknown masses and charge radii of proton-rich nuclei. The theoretical uncertainty of this observable is analyzed in different regions. Finally, we use this robust correlation to predict nuclear binding energies and charge radii based on experimental data from Refs. \cite{AME201,AME202,rc13,rc21}. \footnote{For convenience, in this paper, the database in Refs. \cite{AME201,AME202} is referred to as AME20 (``atomic mass evaluation 2020"); those in Refs. \cite{rc13,rc21} as Rc13 and Rc21 (``2013 and 2021 compilations of charge radii"), respectively.} The validity of our predictions is supported by experimental data from AME20, Rc13, and Rc21, as well as previous predictions from Refs. \cite{DRHBc22,DRHBc24,zong20,zong-self,BAO2024,FU2024,bao71,bao75,bao76,AME202}.
	
	This paper is organized as follows. In Sec. \ref{sec:the}, we propose the correlation involving the new observable, and relate this observable to the two-parameter Fermi model of nuclear charge density. In Sec. \ref{sec:pro}, we analyze the robustness and uncertainty of the estimated observable. In Sec. \ref{sec:ver}, we verify the predictions of nuclear binding energies and charge radii by comparing them with experimental data and predictions from other models. Finally, we summarize this work in Sec. \ref{sec:sum}.
	
	\section{correlation for mirrors}\label{sec:the}
	According to the Bethe-Weizs\"acker formula, the nuclear binding energies of the mirror pair, characterized by proton or neutron number equal to $Z$ or $N$, can be schematically written as 
	\begin{equation}
		\begin{aligned}
			B(Z,N)=&a_v A-a_s A^{2/3}-B_{\rm coul}(Z,N)\\
			&-a_{\rm sym}(\frac{A}{2}-N)^2A^{-1}-a_{\rm pair}\delta_{N,Z}A^{-1/2}\\
			B(N,Z)=&a_v A-a_s A^{2/3}-B_{\rm coul}(N,Z)\\
			&-a_{\rm sym}(\frac{A}{2}-N)^2A^{-1}-a_{\rm pair}\delta_{N,Z}A^{-1/2},\\
		\end{aligned}
	\end{equation}
	where $Z$ and $N$ represent the proton or neutron number of these two mirror nuclei, and $A$ is their mass number. Among the five terms of the Bethe-Weizs\"acker formula, only the Coulomb term $B_{\rm coul}$ differs between mirror nuclei. Thus, the difference in binding energy between the mirror nuclei arises solely from the Coulomb term. The reduced Coulomb displacement energy of mirror nuclei can be written as
	\begin{equation}\label{eq:del-def}
		\begin{aligned}
			\Delta \varepsilon (Z,N)&=\frac{\varepsilon(Z,N)-\varepsilon(N,Z)}{(N-Z)}\\
			&=\frac{B_{\rm coul}(N,Z)-B_{\rm coul}(Z,N)}{A(N-Z)},
		\end{aligned}
	\end{equation}
	where $\varepsilon$ is the binding energy per nucleon as $\varepsilon(Z,N)=\frac{B(Z,N)}{A}$. We further define a new nuclear observable, $\eta(Z,N)$, as
	\begin{equation}\label{eq:eta-def}
		\eta(Z,N)=\Delta \varepsilon (Z,N)\times R_c(Z,N),
	\end{equation} 
	where $R_c(Z,N)$ is the RMS charge radius. The unit of $\eta$ is MeV $\times$ fm $=\frac{1}{197}$ in natural unit system, making $\eta$ a dimensionless observable. If $\eta$ can be determined independently from experiments, Eq. (\ref{eq:eta-def}) directly links two nuclear masses and one charge radius of two mirror nuclei. This creates a long-range yet straightforward correlation with as few as possible nuclei, allowing predictions of nuclear mass and charge radius around proton drip line.
	
	The nuclear charge density $Ze\rho(\vec r)$ should be normalized as follows
	\begin{equation}\label{eq:nor}
		\int \rho(\vec r)d\vec r=1.
	\end{equation}
	Then, 
	\begin{equation}\label{eq:rc-def}
		R_c(Z,N)=\sqrt{\int r^2\rho(\vec r)d\vec r}.
	\end{equation}
	The charge density $Ze\rho(\vec r)$ can also be used to calculate $B_{\rm coul}$. This calculation typically involves decoupling $B_{\rm coul}$ into three components \cite{janecke}: 
	\begin{equation}\label{eq:b-decouple}
		B_{\rm coul}=B_{\rm dir}+B_{\rm exch}+B_{\rm s.o.},
	\end{equation}
	corresponding to the direct Coulomb energy (dir), exchange energy (exch) and spin-orbit coupling (s.o.). Accordingly, we should calculate 
	\begin{equation}\label{eq:eps-decouple}
		\Delta\varepsilon=\Delta\varepsilon_{\rm dir}+\Delta\varepsilon_{\rm exch}+\Delta\varepsilon_{\rm s.o.},
	\end{equation}
	separately. Each term in Eq. (\ref{eq:eps-decouple}) is a functional of $\rho(\vec r)$, making $\eta$ a functional of $\rho(\vec r)$, given that $R_c$ is also a functional of $\rho(\vec r)$ according to Eq. (\ref{eq:rc-def}). By assuming a reasonable charge density $\rho$, one can calculate the $\eta$ value independently of experiments, which may serve the prediction of unknown masses and charge radii. 
	
	In this work, we consider the charge density from the two-parameter Fermi model as 
	\begin{equation}\label{eq:2pf}
		\rho(r)=\frac{\rho_0}{1+\exp\left(\frac{r-R}{a}\right)},
	\end{equation}
	which is an isotropic charge distribution. The parameter $\rho_0$ is introduced to satisfy the normalization condition of Eq. (\ref{eq:nor}). $R$ corresponds to the nuclear geometry radius, where $\rho(R)=\rho_0/2$. $a$ represents the thickness of nuclear surface dispersion. In Sec. \ref{sec:pro}, we will demonstrate that with the two-parameter Fermi model, $\eta$ is a robust observable, and largely unaffected by variations in the parameters $a$ and $R$.
	
	\subsection{RMS charge radius}
	
	With the charge density of Eq. (\ref{eq:2pf}), we evaluate how $R_c$ varies with respect to $a$ and $R$. According to Eq. (\ref{eq:rc-def}) we have
	\begin{equation}
		R_c=\sqrt{\int\frac{\rho_0\times 4\pi r^4dr}{1+\exp\left(\frac{r-R}{a}\right)}},
	\end{equation}
	where
	\begin{equation}
		\rho_0=\frac{1}{\int\frac{4\pi r^2dr}{1+\exp\left(\frac{r-R}{a}\right)}}.
	\end{equation}
	We introduce the function of
	\begin{equation}
		f(t)=\sqrt{\frac{5}{3}\int\frac{x^{4}dx}{1+\exp\left(\frac{x-1}{t}\right) }\left/\int\frac{x^{2}dx}{1+\exp\left(\frac{x-1}{t}\right) }\right.},
	\end{equation}
	where the factor $\frac{5}{3}$ is introduced to insure that $f(0)=1$. Then we have
	\begin{equation}\label{eq:r2-unmod}
		R_c=\sqrt{\frac{3}{5}}Rf\left(\frac{a}{R}\right),
	\end{equation}
	Additionally, as demonstrated by Fig. 3 of Ref. \cite{rc21}, the difference of charge radii between mirror nuclei is also proportional to $(N-Z)/A$ due to isospin asymmetry, supported by theoretical calculations \cite{radius-diff}. To include the effect of isospin asymmetry on charge radii, we modify Eq. (\ref{eq:r2-unmod}) to
    \begin{equation}\label{eq:r2}
        R_c (Z,N)=\sqrt{\frac{3}{5}}Rf\left(\frac{a}{R}\right)+\theta\frac{N-Z}{A},
    \end{equation}
    where $\theta$ is a coefficient to be determined, and $a$ and $R$ parameters are from the nuclei with $N\simeq Z\simeq A/2$.
	
	\subsection{Direct term}\label{sec:dir}
	To obtain $\Delta\varepsilon_{\rm dir}$, one can first calculate the electrical field strength at given $r$ spherical shell. According to the Fermi-model charge density, it should be
	\begin{equation}\label{eq:e}
		E(r)=Z\frac{e}{4\pi\varepsilon_0}\frac{\int_0^r\rho(r^\prime)\times 4\pi r^{\prime 2}dr^\prime}{r^2}.
	\end{equation}
	Then, the direct term of Coulomb energy of a nucleus with proton number $Z$ reads
	\begin{equation}\label{eq:b-dir}
		\begin{aligned}
			B_{\rm dir}=&\frac{1}{2}\varepsilon_0\int E^2(r)\times 4\pi r^2dr\\
			=&\frac{3}{5}Z^2\frac{e^2}{4\pi\varepsilon_0}\frac{1}{R}h\left(\frac{a}{R}\right),
		\end{aligned}
	\end{equation} 
	where
	\begin{equation}
		h(t)=\frac{5}{6}\displaystyle\int \frac{
			\left[
			\int^x_0\frac{y^2dy}{1+\exp\left(\frac{y-1}{t}\right) }
			\left/\int\frac{x^{2}dx}{1+\exp\left(\frac{x-1}{t}\right) }\right.
			\right]^2
		}{x^2}dx.
	\end{equation}
	The factor $\frac{5}{6}$ is introduced so that $h(0)=1$. Assuming two mirror nuclei share the same $a$ and $R$ parameters for their charge density, then the contribution from the direct term to the reduced Coulomb displacement energy reads
	\begin{equation}
		\Delta\varepsilon_{\rm dir}=\frac{3}{5}\frac{e^2}{4\pi\varepsilon_0}\frac{1}{R}h\left(\frac{a}{R}\right).
	\end{equation}
	
	Correspondingly, the contribution from direct term to $\eta$ observable would be 
	\begin{equation}
		\eta_{\rm dir}=\frac{3}{5}\sqrt{\frac{3}{5}}\frac{e^2}{4\pi\varepsilon_0}\left[f\left(\frac{a}{R}\right)+\theta\frac{N-Z}{\sqrt{\frac{3}{5}}AR}\right]h\left(\frac{a}{R}\right).
	\end{equation}
	Assuming $a=0$, i.e., with a uniform spherical charge distribution, and $\theta=0$ in Eq. (\ref{eq:r2}), i.e., without isospin asymmetry, $\eta_{\rm dir}$ reduces to a non-dimensional constant, $\frac{3}{5}\sqrt{\frac{3}{5}}\frac{e^2}{4\pi\varepsilon_0}$. For convenience, we define 
	\begin{equation}
		\eta_0=\frac{3}{5}\sqrt{\frac{3}{5}}\frac{e^2}{4\pi\varepsilon_0},
	\end{equation}
	as the unit for the $\eta$ evaluation. Then, we have
	\begin{equation}
		\eta_{\rm dir}=\eta_0\left[f\left(\frac{a}{R}\right)+\theta\frac{N-Z}{\sqrt{\frac{3}{5}}AR}\right]h\left(\frac{a}{R}\right).
	\end{equation}
	
	\subsection{Exchange term}\label{sec:exch}
	Based on the Fermi-gas model, Bethe and Bacher \cite{bethe,janecke22} determined the exchange term of the Coulomb energy as
	\begin{equation}
		B_{\rm exch}=-\frac{3}{2}\left(\frac{3}{8\pi}\right)^{\frac{1}{3}}\frac{Z^{4/3}e^2}{4\pi\varepsilon_0}\int \left[\rho(r)\right]^{4/3}4\pi r^2dr,
	\end{equation}
	We introduce
	\begin{equation}
		k(t)=3^{-\frac{1}{3}}
		\int \frac{x^2dx}{\left[1+e^{\frac{x-1}{t}}\right]^{\frac{4}{3}}}
		\left/
		\left[\int \frac{x^2dx}{1+e^{\frac{x-1}{t}}}\right]^{\frac{4}{3}}\right.,
	\end{equation}
	where the factor $3^{-\frac{1}{3}}$ is introduced so that $k(0)=1$. Then we have
	\begin{equation}
		B_{\rm exch}=-\frac{3}{4}\left(\frac{3}{2\pi}\right)^{\frac{2}{3}}\frac{e^2}{4\pi\varepsilon_0}\frac{Z^{\frac{4}{3}}}{R}k\left(\frac{a}{R}\right).
	\end{equation}
	The contribution of exchange term to $\Delta\varepsilon$ is then written as
	\begin{equation}\label{eq:eta-exch}
		\begin{aligned}
			\Delta\varepsilon_{\rm exch}=&-\frac{3}{4}\left(\frac{3}{2\pi}\right)^{\frac{2}{3}}\frac{e^2}{4\pi\varepsilon_0}\frac{N^{\frac{4}{3}}-Z^{\frac{4}{3}}}{A(N-Z)R}k\left(\frac{a}{R}\right)\\
			\simeq &\frac{1}{2}\left(\frac{3}{\pi}\right)^{\frac{2}{3}}\frac{e^2}{4\pi\varepsilon_0}\frac{1}{A^{\frac{2}{3}}R}k\left(\frac{a}{R}\right),
		\end{aligned}
	\end{equation}
	where $\frac{(N^{4/3}-Z^{4/3})}{(N-Z)}\simeq\frac{4}{3}\left(\frac{N+Z}{2}\right)^{\frac{1}{3}}$ is adopted, according to differential mean value theorem.
	
	\subsection{Spin-orbit coupling term}\label{sec:so}
	According to Eq. (42) of Ref. \cite{janecke}, the reduced Coulomb displacement contribution from the spin-orbit coupling is written as
	\begin{equation}\label{eq:so}
		\begin{aligned}
			\Delta\varepsilon_{\rm s.o.}=&-0.6\frac{1}{2m_\pi^2}e\frac{g_\pi-g_\nu-1}{A}\\
			&\times\int\rho(r)\frac{1}{r}\frac{dV_{\rm core}}{dr}4\pi r^2dr,
		\end{aligned}
	\end{equation}
	where the $g$ factors for proton and neutron are $g_\pi=5.586$ and $g_\nu=-3.826$, respectively, and the proton mass is taken as $m_\pi=938$ MeV. To obtain Eq. (\ref{eq:so}), it is assumed that the spin-orbit coupling sum of $S=\sum \vec l\cdot\vec \sigma$ is on average proportional to nucleon number with slope of 0.6, as demonstrated in Fig. 6 of Ref. \cite{janecke}.
	
	Given that charge density of the Fermi model is isotropic, the gradient of electric field $\frac{dV_{\rm core}}{dr}=E(r)$, where $E(r)$ is the electric field strength as defined in Eq. (\ref{eq:e}). Using Gauss's Law for a static electric field, $E(r)\times 4\pi r^2=\frac{Q(r)}{\varepsilon_0}$, where $Q(r)$ is the electric charge within a sphere of radius $r$, and thus $Q(r)=Ze\int_0^r\rho(r^\prime)4\pi r^{\prime 2}dr^\prime$. Note that $\Delta\varepsilon_{\rm s.o.}$ actually involves two nuclei with different proton numbers. Thus, we introduce the average $Q(r)$ for two mirror nuclei into the spin-orbit coupling term as
	\begin{equation}
		\overline  Q(r)=\frac{N+Z}{2}e\int_0^r\rho(r^\prime)4\pi r^{\prime 2}dr^\prime.
	\end{equation}
	Then, we have
	\begin{equation}\label{eq:delta-so}
		\begin{aligned}
			\Delta\varepsilon_{\rm s.o.}=&-0.6\frac{1}{2m_\pi^2}\frac{e}{\varepsilon_0}\frac{g_\pi-g_\nu-1}{A}\int\rho(r)\frac{1}{r}\overline Q(r)dr\\
			=&-0.3\frac{1}{2m_\pi^2} \frac{e^2}{\varepsilon_0}(g_\pi-g_\nu-1)\\
			&\times\int\rho(r)\frac{1}{r}\left\{\int_0^r\rho(r^\prime)4\pi r^{\prime 2}dr^\prime\right\}dr\\ 
			=&-0.3\frac{1}{2m_\pi^2} \frac{e^2}{4\pi\varepsilon_0}(g_\pi-g_\nu-1)\frac{1}{R^3}g\left(\frac{a}{R}\right),\\ 
		\end{aligned}
	\end{equation}
	where 
	\begin{equation}
		g(t)=
		\displaystyle\int
		\frac{\int^x_0 \frac{y^2}{1+e^{\frac{y-1}{t}}}dy}
		{x\left\{1+e^{\frac{x-1}{t}}\right\}}dx
		\left/
		\left[\int \frac{x^2dx}{1+e^{\frac{x-1}{t}}}\right]^2
		\right..
	\end{equation}
	One sees that $g(0)=1$.
	
	\subsection{Total $\eta$}\label{sec:eta}
	
	In Secs. \ref{sec:dir}, \ref{sec:exch} and \ref{sec:so}, we assume that the mirror nuclei share $a$ and $R$ parameters. However, normally it's not the case. Thus, we believe that the values of $a$ and $R$ should be taken as the average between two mirror nuclei, or according to differential mean value theorem, they should be from the nucleus located midway between the two mirrors. Thus, in the calculation of $\Delta\varepsilon (Z,N)$ using Eqs. (\ref{eq:b-dir}), (\ref{eq:eta-exch}), and (\ref{eq:delta-so}), the $a$ and $R$ parameters should be taken from nucleus with $N\simeq Z\simeq A/2$. Additionally, according to Eq. \ref{eq:r2}, we determine the charge radius $R_c(Z,N)$ with the $a$ and $R$ parameters of the nucleus with $N\simeq Z\simeq A/2$. Thus, the total $\eta$ with contributions from all the three term can be written as 
	\begin{widetext}
		\begin{equation}\label{eq:eta-all}
				\eta= \eta_0\left[f\left(\frac{a}{R}\right)+\theta\frac{N-Z}{\sqrt{\frac{3}{5}}AR}\right]\left\{h\left(\frac{a}{R}\right)-\frac{5}{12}\left(\frac{3}{\pi}\right)^{\frac{2}{3}}A^{-\frac{2}{3}}k\left(\frac{a}{R}\right)-\frac{g_\pi-g_\nu-1}{4m_\pi^2 R^2}g\left(\frac{a}{R}\right)\right\}.
		\end{equation}
	\end{widetext}
	
	We note that the factor $\frac{5}{6}\left(\frac{3}{\pi}\right)^{\frac{2}{3}}A^{-\frac{2}{3}}$ approximately ranges from  0.17 to 0.03, and the factor $\frac{g_\pi-g_\nu-1}{4m_\pi^2 R^2}$ from 0.014 to 0.003 for $A$ from  10 to 130 with widely adopted $R\sim 1.2A^{1/3}$ rule. Since the functions $h,k,g$ are all in the same order, with $h(0)=k(0)=g(0)=1$. It can be seen that the direct term contributes most to $\eta$, the exchange term contributes less by an order of magnitude, and the spin-orbit coupling term contributes even less by another order of magnitude. We also note that the contributions of exchange term and the spin-orbit coupling term are related to the mass number by a factor of $A^{-\frac{2}{3}}$, considering that the nuclear geometry radius $R\propto A^{\frac{1}{3}}$. Therefore, we expect these two terms to become even less important as the mass number $A$ increases.
	
	\section{property of the correlation}\label{sec:pro}
		According to Eq. (\ref{eq:eta-def}), the proposed correlation for mirror nuclei is governed by the observable $\eta$. Therefore, the properties of this correlation are largely determined by this observable. Using Eq. (\ref{eq:eta-all}), we can discuss the robustness of $\eta$ and its uncertainty arising from our modeling.
	\subsection{Robustness}
	
	The robustness of $\eta$ refers to its stability when varying the $a$ or $R$ parameters. Such robustness makes our prediction based on the correlation of Eq. (\ref{eq:eta-def}) largely immune to the uncertainties of $a$ and $R$ parameter, as well as other potential local structural irregularities, such as shell effects and pairing.
	
	\begin{figure}
	\includegraphics[angle=0,width=0.45\textwidth]{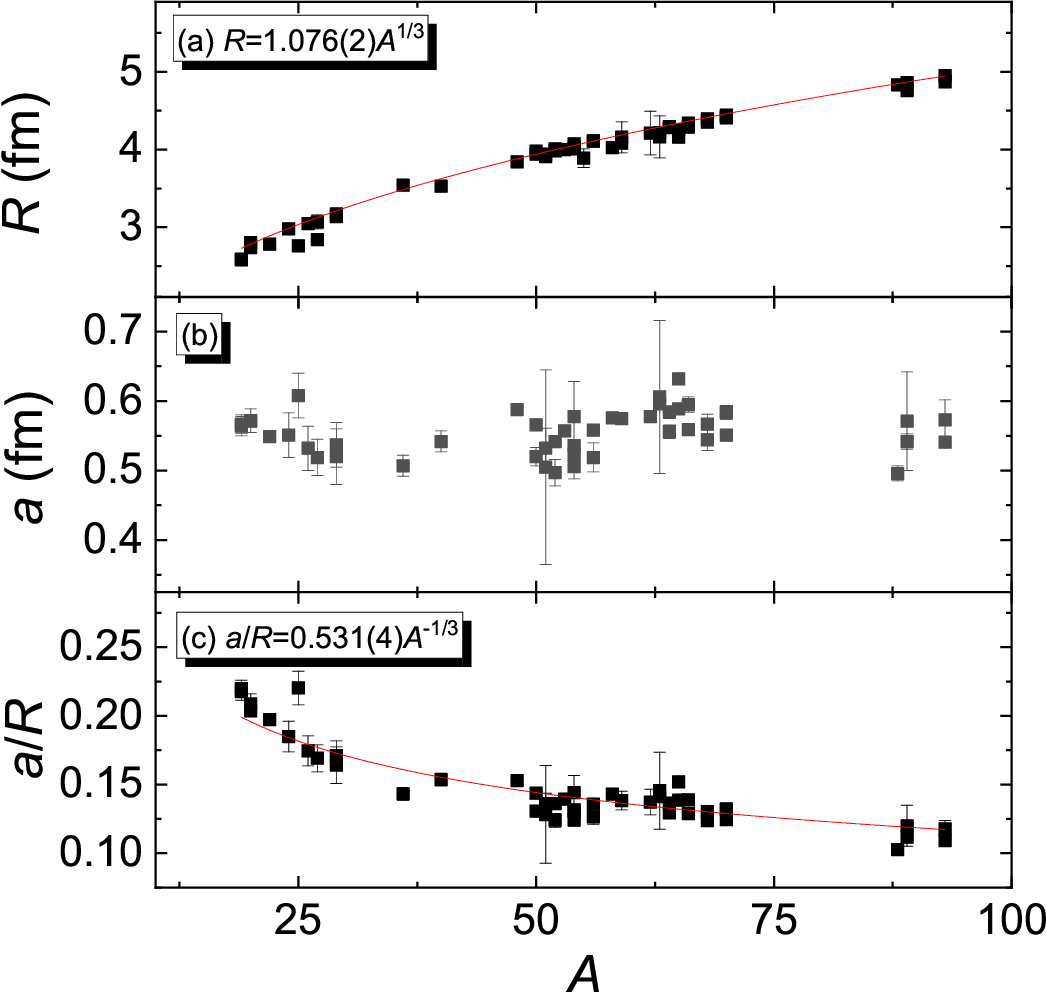}
	\caption{(Color online)
		$a$ and $R$ for $A<100$ from experimental compilation \cite{aRdata}. $R$ is related to $A$ as expected by the $A^{1/3}$ law, and the linear fit with zero intercept gives $R\simeq 1.076(2) A^{1/3}$ fm. $a$ appears to be independent of $A$, but the $a/R$ ratio generally follows the regularity of $a/R=0.531(4)A^{-1/3}$. The red curves represent the results of the fitting.}\label{fig:aR_A}
	\end{figure} 
	
	Thus, we should first observe how $a$ and $R$ parameters vary to understand the robustness of $\eta$. Based on the compilation of $a$ and $R$ values from experimental data \cite{aRdata}, we plot all the $a$ and $R$ values for $A<100$ in Fig. \ref{fig:aR_A}. In Fig. \ref{fig:aR_A}(a), $R$ appears to be a monotonically increasing function of $A$. A linear fit between $R$ and $A^{1/3}$ is carried out, suggesting $R=1.076(2)A^{1/3}$ fm. This indicates that the $R$ parameter well follows the commonly accepted $A^{1/3}$ law of nuclear size, further confirming the nature of $R$ as the nuclear geometry radius. In Fig. \ref{fig:aR_A}(b), the $a$ parameter varies between 0.5 and 0.6 fm, with no clear systematics observed. However, we note that in Eq. (\ref{eq:eta-all}), the $a$ parameter affects $\eta$ evaluation only through the $a/R$ ratio. Thus, only the $a/R$ ratio matters. We plot $a/R$ ratio in Fig. \ref{fig:aR_A}(c), and fit it to a proportional relation of $a/R\propto A^{-1/3}$. Most $a/R$ ratios are found close to $a/R=0.531(4)A^{-1/3}$, suggesting that from the perspective of $a/R$ ratio, the variation in $a$ is much smaller than that in $R$, and behaves nearly as a constant. Therefore, although the $a$ parameter seems to vary randomly, there should be some systematic evolution of $\eta$ with the $R$ and $a/R$ ratio, i.e. the mass number $A$. Moreover, as $A$ increases, $a/R$ ratio approaches zero following the $A^{-1/3}$ law. This suggests that for heavier nuclei, the charge distribution becomes similar to a uniformly charged sphere, and $\eta$ should approach to $1\eta_0$ for $N\sim Z$ nuclei.
	
\begin{figure}
		\includegraphics[angle=0,width=0.45\textwidth]{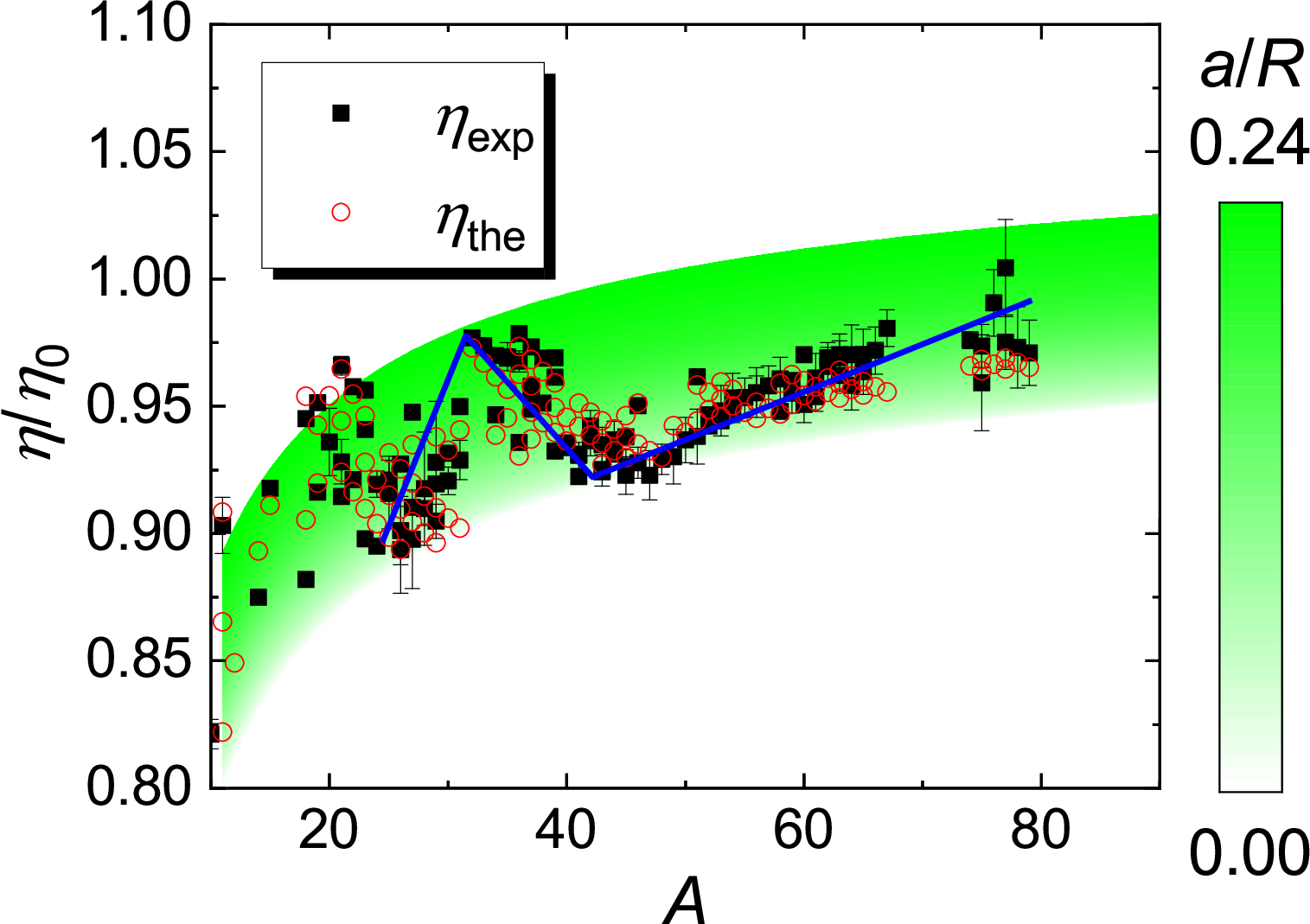}
		\caption{(Color online)
			$\eta$ against mass number $A$. The green color mapping of the $a/R$ ratio illustrates how $\eta$, as calculated from Eq. (\ref{eq:eta-all}), varies with the $a/R$ ratio and the $R$ parameter, using the commonly adopted $R=1.2A^{1/3}$ fm and assuming perfect isospin conservation with $\theta=0$. 			
			The black square ($\blacksquare$) represents the experiential $\eta_{\rm exp}$, calculated using Eq. (\ref{eq:eta-def}), with the binding energy and RMS charge radii data taken from the Rc13, Rc21, and AME20 databases. The blue polyline schematically highlights the inflection points of the $\eta_{\rm exp}$ evolution across different mass regions. The red circle ($\color{red}\circ$) represents the theoretical $\eta_{\rm the}$, calculated from Eq. (\ref{eq:eta-all}), using the best-fit parameters $R=1.088A^{1/3}$, $a/R=0.584A^{-1/3}$ and $\theta=1.047$.
		}\label{fig:eta-A-exp}
	\end{figure}

	Using Eq. (\ref{eq:eta-all}) with numerical integration algorithms, we can observe the general tendency of $\eta$ as a function of the mass number $A$, i.e., the $R$ parameter following the commonly adopted $R=1.2A^{1/3}$ regularity, and the color-mapped $a/R$ ratio up to 0.24, with perfect isospin conservation ($\theta=0$), as shown in Fig. \ref{fig:eta-A-exp}. The value 0.24 is the upper boundary of the $a/R$ ratio, as shown in Fig. \ref{fig:aR_A}(c). It can be observed that $\eta$ only varies within a range of approximately $0.1\eta_0$ for all possible $a/R$ ratios and $R$ values (mass number $A$). Therefore, the correlation between binding energy difference and charge radius of mirror nuclei, governed by the $\eta$ observable, should be robust regardless of the mass region (the $R$ parameter), and the numerical fluctuation in the $a/R$ ratio, as long as the $R\propto A^{1/3}$ law holds. 
	
	Using the experimental binding energy and RMS charge radii data from the AME20, Rc13, and Rc21 databases, 
	{\color{blue}
		as well as Ref. \cite{Sc-self,Sc-6,Sc-8,K-self}
	}, 
	we calculate $\eta$ for all available mirror nuclei according to Eq. (\ref{eq:eta-def}). We plot the experimentally determined $\eta_{\rm exp}$ as a function of mass number $A$ in Fig. \ref{fig:eta-A-exp}. Most of the $\eta_{\rm exp}$ values fall within the expected $\eta$ trend, as indicated by the green color mapping of the $a/R$ ratio, except for few light nuclei around $A\sim 20$ or $A=10$. We also note that the experimental error of $\eta_{\rm exp}$ is smaller than the uncertainty in $\eta$ due to fluctuations in the $a/R$ ratio between 0 and 0.24, as demonstrated by the green color mapping in Fig. \ref{fig:eta-A-exp}. Therefore, fluctuations in the $a/R$ ratio may still introduce significant modeling uncertainty to our following predictions based on Eq. (\ref{eq:eta-all}), compared to the experimental error.
	
	To make predictions, $\eta$ must be calculated using Eq. (\ref{eq:eta-all}) independently of experimental data. This requires explicit relations between mass number $A$ and the $R$ parameter, as well as between $A$ and the $a/R$ ratio, for $N\simeq Z$ nuclei. However, the $a$ and $R$ data compilation \cite{aRdata} includes only $Z<N$ nuclei. Therefore, we cannot directly apply the $R$ and $a/R$ regularities in Fig. \ref{fig:aR_A}(a) and (c) to calculate $\eta$. However, we assume that $R\propto A^{1/3}$ and $a/R\propto A^{-1/3}$ laws hold for $N\simeq Z$ nuclei as well. Thus, we express $R$ as $R=r_0 A^{1/3}$ and $a/R$ as $a/R=(a/R)_0 A^{-1/3}$ in Eq. (\ref{eq:eta-all}), and fit this equation to experimental $\eta$ values using simplex method \cite{simplex}, treating $r_0$, $(a/R)_0$ and $\theta$ as fitting parameters. The fit yields $r_0=1.088$ fm, $(a/R)_0=0.584$, and $\theta=1.047$ fm. These fitting results are numerically close to the regularities $R=1.076(2)A^{1/3}$ and $a/R=0.531(4)A^{-1/3}$ obtained from fitting the $a$ and $R$ data compilation \cite{aRdata}, as shown in Fig. \ref{fig:aR_A}. Furthermore, we apply the parameters $r_0=1.088$ fm, $(a/R)_0=0.584$, and $\theta=1.047$ fm to Eq. (\ref{eq:eta-all}), and plot theoretically calculated $\eta_{\rm the}$ in Fig. \ref{fig:eta-A-exp}. The agreement between experimental and calculated values is evident within an RMS deviation (RMSD) of $\sim 0.05\eta_0$. Therefore, we trust the validity of our fitting results and use them for further uncertainty analysis and predictions.  
	
	\subsection{Uncertainty}\label{sec:unc}
	 
	\begin{figure}
		\includegraphics[angle=0,width=0.45\textwidth]{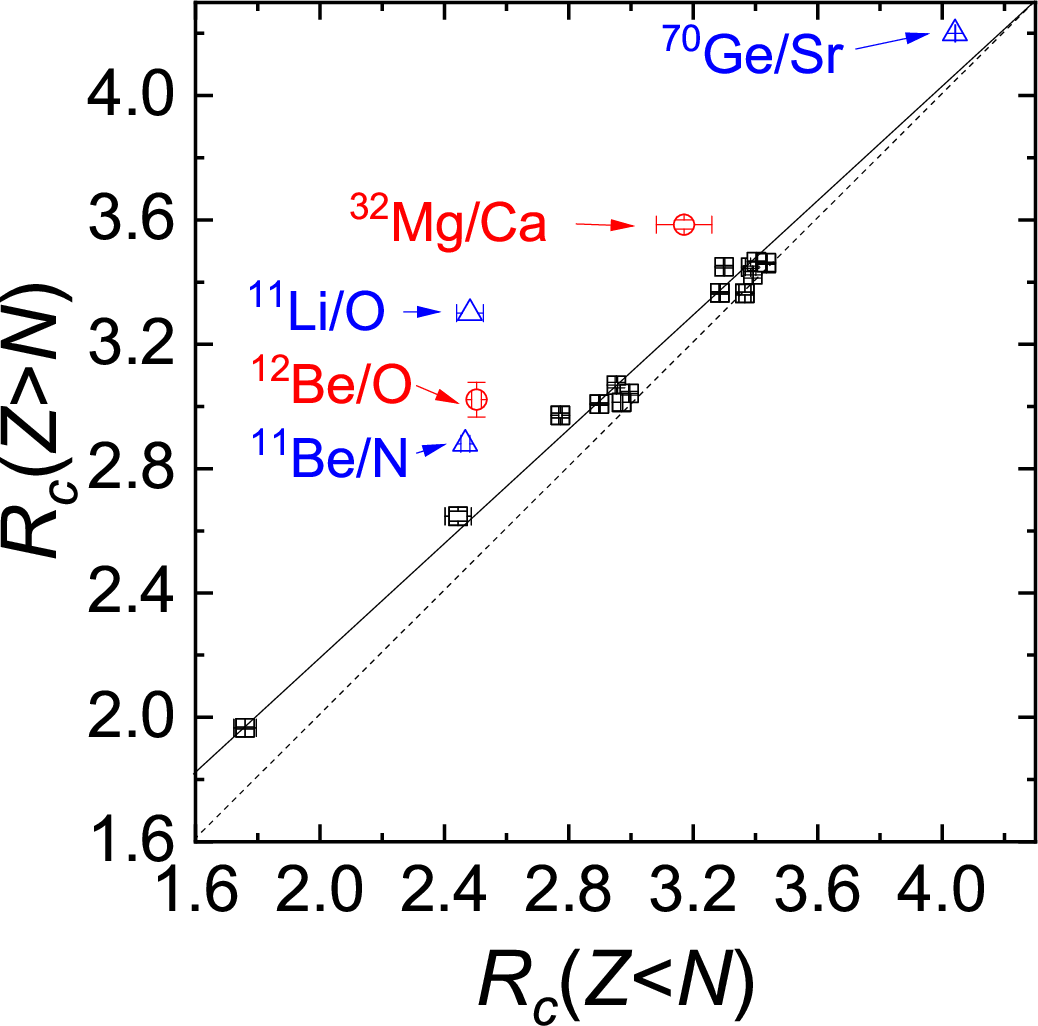}
		\caption{(Color online)
			$R_c$ of mirror nuclei against each other. The diagram is designed in square shape, so that the diagonal dash line corresponds to perfect symmetry of $R_c$ in mirror nuclei, and possibly symmetry of charge density shape. The black square ($\square$) corresponds to $R_c$ of mirror nuclei from experimental $R_c$ compilation of Rc13 and Rc21. The solid line schematically presents the systematics of the experimental correlation between $R_c$ of mirror nuclei. 			
			$R_c(Z<N)$ of red-circle ($\color{red}\circ$) data are from Rc13/21, while their $R_c(Z>N)$ are from Refs. \cite{DRHBc22,DRHBc24}, because those experimental $R_c(Z>N)$ values are still unavailable. The blue-triangles ($\color{blue}\vartriangle$) are determined by $R_c(Z<N)$ from Rc13/21, and $R_c(Z>N)$ from our own DRHBc calculations, given they are not listed in Refs. \cite{DRHBc22,DRHBc24}. The uncertainties of calculated $R_c(Z>N)$ are not provided.
			The red circle and the blue triangle data are all off the experimental systematics.
		}\label{fig:sym-rc}
	\end{figure} 
	
	The uncertainty of the $\eta$ estimation is crucial for making reliable predictions on binding energies and charge radii. We attribute most of the modeling uncertainty to the charge-density asymmetry of mirror nuclei. In the derivation of Sec. \ref{sec:the}, we assume that the mirror nuclei share the same charge density distribution function, $\rho(r)$. This means that $R$ and $a/R$ are the same for both mirror nuclei. However, this assumption is unlikely to be precisely accurate, and cannot be verified using the data in Ref. \cite{aRdata}, which only includes $a$ and $R$ from $Z<N$ nuclei. Nevertheless, we can compare $R_c$s of mirror nuclei, which may indicate the $R$ and $a/R$ asymmetry, according to Eq. (\ref{eq:r2-unmod}). In Fig. \ref{fig:sym-rc}, we plot the experimental $R_c$ values of mirror nuclei against each other in a square diagram, where the diagonal line represents perfect $R_c$ symmetry. All the the experimental data points lie close to the diagonal line, indicating that mirror nuclei do have nearly identical $R_c$, and therefore possibly similar $R$ parameter and $a/R$ ratio. However, we also observe that $R_c$ values for $Z>N$ nuclei are systematically, though slightly, larger than those of their $N>Z$ counterparts, reflecting a slight charge density asymmetry. We also note that as $R_c$ increases, the experimental data points systematics and gradually approach the diagonal, suggesting that the asymmetry is reduced in heavier nuclei with larger $R_c$. This asymmetry was also observed in Refs. \cite{rc21,radius-diff}, where the difference of mirror $R_c$ values was related to $(N-Z)/A$. The modeling uncertainty in the estimated $\eta$ due to charge-density asymmetry is expected to decrease for larger mass number $A$. Therefore, we should quantify the uncertainty in $\eta$, separately for light and heavier nuclei.
	
	Furthermore, by examining Fig. \ref{fig:eta-A-exp}, we note that the experimental $\eta_{\rm exp}$ exhibits several inflection points, which divide the data into four mass regions, each with a distinct trend in the evolution of $\eta_{\rm exp}$. In the mass range $10<A\leq 24$, $\eta_{\rm exp}$ appears more random. As shown by the blue polyline of Fig. \ref{fig:eta-A-exp}, in the mass ranges $24<A\leq 31$ and $41<A$ region, $\eta_{\rm exp}$ generally increases monotonically with $A$, whereas in the range $31<A\leq 41$, it decreases. Empirically, the uncertainty of estimated $\eta_{\rm the}$ differs in the mass regions with different $\eta_{\rm exp}$ trends. Therefore, we should quantify the uncertainty in $\eta_{\rm the}$ separately for each of these four regions.
	
	In each mass region, we estimate the modeling uncertainty of our $\eta_{\rm the}$ calculation, using the maximum-likelihood method from Refs. \cite{uncertainty-method,FU2024}, i.e., the following iterative process  
	\begin{equation}\label{eq:eta-unc}
		\begin{aligned}
			\sigma^{2 }_{\eta}&=\frac{\sum w_i^2\left[\left(\eta_{\rm exp}^i-\eta_{\rm the}^i-\mu\right)^2-(\sigma_{\rm exp}^i)^2\right]}{\sum w_i^2}  \\
			\mu&=\frac{\sum w_i\left[\eta_{\rm exp}^i-\eta_{\rm the}^i\right]}{\sum w_i}\\
			w_i&=\frac{1}{(\sigma^i_{exp })^2+\sigma^2_{\eta}}.
		\end{aligned}
	\end{equation}
	Here, $\eta^i_{\rm exp}$, $\eta^i_{\rm the}$ and $\sigma^i_{\rm exp}$ represent the $i$-th experimental and theoretical $\eta$ values, and the corresponding experimental error, in the region under investigation, respectively. $\sigma_{\eta}$ represents the modeling uncertainty in our $\eta_{\rm the}$ calculation, and $\mu$ denotes the systematic deviation of $\eta_{\rm the}$ from experiments. The calculated values of $\sigma_{\eta}$ and $\mu$ for the four regions $10<A\leq 24$, $24<A\leq 31$, $31<A\leq 41$ and $41<A$ are presented in Table \ref{tab:eta-unc}.
	
	\begin{table}
		\caption{$\sigma_{\eta}$ and $\mu$, representing the modeling uncertainty and systematic deviation of our calculated $\eta_{\rm the}$, obtained from the iterations using Eq. (\ref{eq:eta-unc}) for four different mass regions.}\label{tab:eta-unc}
		\begin{tabular}{cccccccccccccccccccccccccc}
			\hline\hline
			 &$\mu$&$\sigma$ & &$\mu$& $\sigma$\\
			 \hline
			$10<A\leq 24$ & -0.009&0.024 & $24<A\leq 31$ & 0.009&0.014 \\
			
			$31<A\leq 41$ & 0.001&0.010 &	$41<A$  &0.003& 0.009 \\
			\hline\hline
		\end{tabular}
	\end{table}
	
	As shown in Table \ref{tab:eta-unc}, the theoretical uncertainty $\sigma_{\eta}$ decreases significantly from 0.024 to 0.009. This confirms that our $\eta_{\rm the}$ estimation introduces  larger uncertainty for lighter nuclei, as anticipated in our previous analysis of Sec. \ref{sec:unc}. Therefore, we accept the $\sigma_{\eta}$ values listed in Table \ref{tab:eta-unc} as the modeling uncertainty of $\eta_{\rm the}$, and we will use these $\sigma_{\eta}$ values in our subsequent analysis and predictions of binding energies and charge radii. 
	
	Moreover, $\mu$ represents the systematic deviation of our $\eta_{\rm the}$ estimation from experimental values. To account for this deviation, we adopt 	
	\begin{equation}\label{eq:z-dep}
		\tilde\eta=\eta_{\rm the}+\mu,
	\end{equation} 
	as our final and optimal estimation of $\eta$.
	
	\section{verification and prediction}\label{sec:ver}
	
	\subsection{Binding energy per nucleon}
	
	According to Eqs. (\ref{eq:del-def}) and (\ref{eq:eta-def}), if the experimental binding energy per nucleon and charge radius of a given nucleus nucleus are, the binding energy per nucleon of its mirror nucleus can be predicted as follows
	\begin{equation}\label{eq:ba-pre}
		\varepsilon^{\rm pre}(N,Z)=\varepsilon^{\rm exp}(Z,N)-\frac{\tilde\eta(Z,N)}{R^{\rm exp}_c(Z,N)}(N-Z).
	\end{equation}
	Here, $\varepsilon^{\rm exp}(Z,N)$ is the experimental binding energy per nucleon of the nucleus with $Z$ protons and $N$ neutrons, and $R^{\rm exp}_c(Z,N)$ is the experimental RMS charge radius of this nucleus. $\tilde\eta$ is calculated with Eq. (\ref{eq:z-dep}), and $\varepsilon^{\rm pre}(N,Z)$ represents the predicted binding energy per nucleon of the mirror nucleus. The prediction uncertainty has two components. The first arises from the experimental errors in $\varepsilon^{\rm exp}(Z,N)$ and $R^{\rm exp}_c$, and is given by
	\begin{equation}\label{eq:ba-err}
		\delta_{\varepsilon(N,Z)}=\sqrt{\delta^2_{\varepsilon(Z,N)}+\frac{\tilde\eta^2(Z,N)(N-Z)^2}{\left[R^{\rm exp}_c(Z,N)\right]^4}\delta^2_{R_c(Z,N)}},
	\end{equation}
	where $\delta_{\varepsilon(Z,N)}$ and $\delta_{R_c(Z,N)}$ are the experimental errors of $\varepsilon^{\rm exp}(Z,N)$ and $R^{\rm exp}_c(Z,N)$, respectively. The second component is due to the modeling uncertainty in our $\tilde\eta$ estimation, expressed as
	\begin{equation}\label{eq:ba-unc}
		\sigma_{\varepsilon(N,Z)}=\frac{|N-Z|}{R^{\rm exp}_c(Z,N)}\sigma_{\tilde\eta}(Z,N).
	\end{equation}
	The total uncertainty is then calculated as $\sqrt{\delta^2_{\varepsilon}+\sigma^2_{\varepsilon}}$. 
	In this work, we predict the binding energies per nucleon of 197 nuclei, which are listed along with their corresponding total uncertainties in the Supplemental Materials of this paper \cite{supp}.

	\begin{figure}
	\includegraphics[angle=0,width=0.45\textwidth]{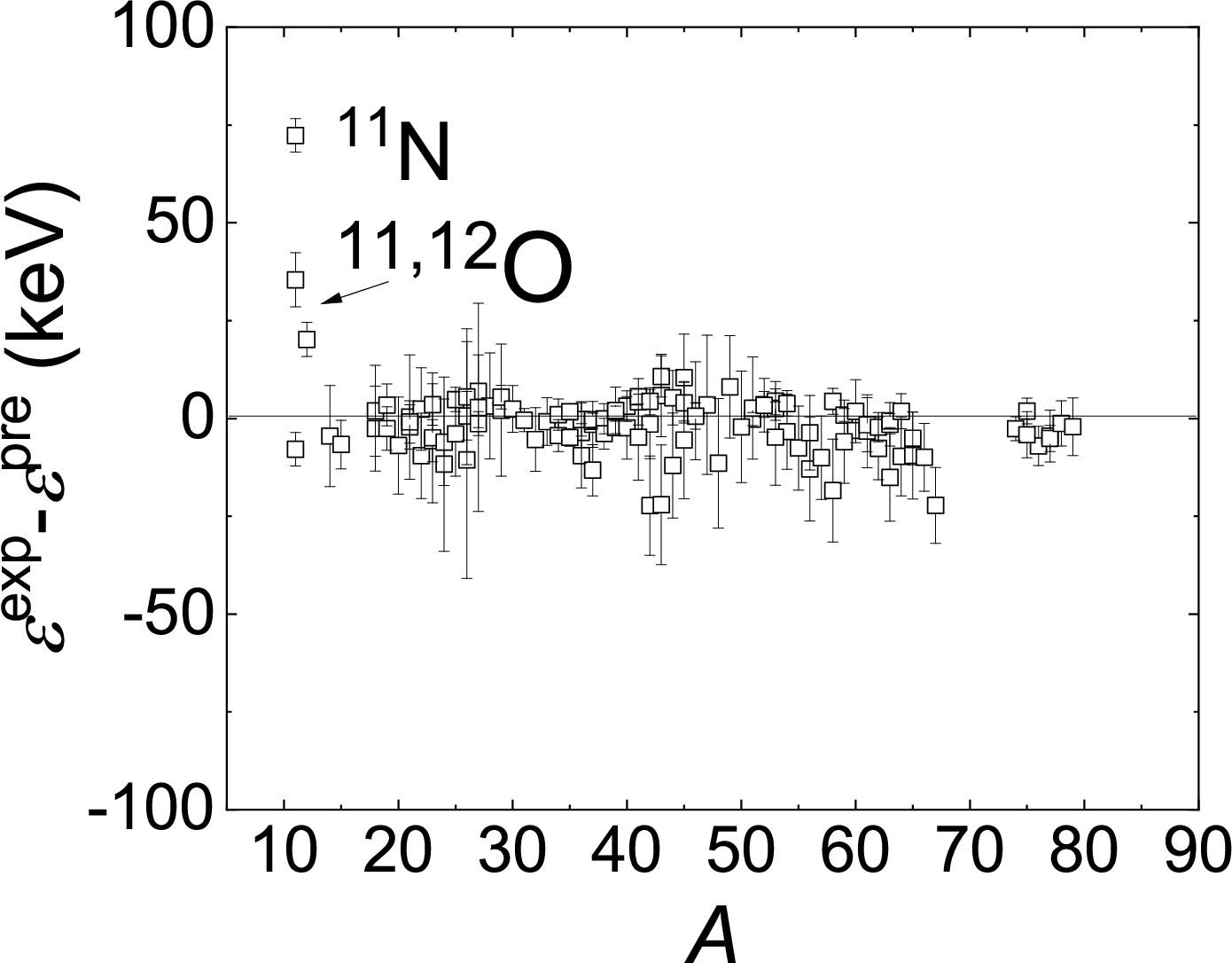}
		\caption{
			Difference between ``predicted" binding energy per nucleon ($\varepsilon^{\rm pre}$) and the experimental values ($\varepsilon^{\rm exp}$) against mass number $A$, for nuclei available in AME20. The prediction uses experimental data from AME20, Rc13 and Rc21, based on Eq. (\ref{eq:ba-pre}). The prediction uncertainty are calculated as $\sqrt{\delta^2_\varepsilon+\sigma^2_\varepsilon}$ using Eq. (\ref{eq:ba-err}) and (\ref{eq:ba-unc}). The obvious inconsistency for $^{11,12}$O is highlighted.
		}\label{fig:ba20to20}
	\end{figure}
	
	To verify our prediction method for binding energy per nucleon $\varepsilon$, we use data from AME20, Rc13 and Rc21 to ``predict" $\varepsilon^{\rm pre}$ values for nuclei, whose $\varepsilon^{\rm exp}$ values are also available in AME20. We present the difference between $\varepsilon^{\rm exp}$ and $\varepsilon^{\rm pre}$ in Fig. \ref{fig:ba20to20}. In most cases, the differences fall within prediction uncertainty, except for $^{11}$N and $^{11,12}$O. Our predicted binding energies for $^{11}$N and $^{11,12}$O obviously differ from experimental values, exceeding the $1\sigma$ confidence interval. The RMSD of the mass excesses is 193 keV, and excluding the clear discrepancies for $^{11}$N and $^{11,12}$O, it is 160 keV. This RMSD is not competitive, compared to other mass predictions based on the mass correlations between mirror nuclei, such as those in Refs. \cite{zong20,BAO2024,FU2024}. However, it is noteworthy that our prediction does not exclude unbound nuclei. For unbound nuclei, Ref. \cite{zong-self} also reported predictions with an RMSD of 183 keV, or 164 keV without $^{11}$Li/O mirror, which is consistent with the performance of our mass predictions. Thus, the inaccuracy of our mass predictions may be attributed to the inclusion of unbound nuclei. 
	
	One question arises: does such a large RMSD aligns with our model uncertainty, or does it stems from a systematic deviation of our predictions from experiments. The $\chi^2$ analysis could provide answers. We define the reduced $\chi^2$ for mass excess as
	\begin{equation}
		\chi^2_M=\frac{1}{N-3}\sum\frac{(M^i_{\rm exp}-M^i_{\rm pre})^2}{\left(\sigma^i_{\rm exp}\right)^2+\left(\sigma^i_{\rm pre}\right)^2},
	\end{equation}
	where $M^i_{\rm exp}$ is the experimental mass excess of the $i$-th nucleus. $M^i_{\rm pre}$, $\sigma^i_{\rm exp}$ and $\sigma^i_{\rm pre}$ are the corresponding predicted value, experimental error, and prediction uncertainty calculated as $\sqrt{\delta^2_\varepsilon+\sigma^2_\varepsilon}$ using Eqs. (\ref{eq:ba-err}) and (\ref{eq:ba-unc}), respectively. The factor $\frac{1}{N-3}$ is introduced instead of $\frac{1}{N}$, because there are three fitting parameters used to fit Eq. \ref{eq:eta-all} to experiments. If $\chi^2_M$ is much larger than 1, then our prediction likely has a systematic deviation from experiments, which could be corrected by introducing a simple phenomenological adjustment to our mass prediction. Conversely, if it is much smaller than 1, there may be an overfitting problem.	
	
	For the entire set of mass excess, $\chi^2_M=1.1$. Excluding $^{11}$N and $^{11,12}$O, $\chi^2_M=0.9$. Both values are close to 1, suggesting that the estimated modeling uncertainty aligns with the RMSD, and that no systematic deviation exists in our prediction. Therefore, no simple correction can be made to further improve our mass prediction.
	
	The disagreements for $^{11}$N and $^{11,12}$O are understandable. As demonstrated in Fig. \ref{fig:sym-rc}, the experimental charge radii of mirror nuclei appear to follow linear systematics, indicating a slight charge-density asymmetry between mirror nuclei. We further plot the charge radii of $^{11}$Li/O in Fig. \ref{fig:sym-rc}, with $R_c=2.48(4)$ for $^{11}$Li from Rc13 and $R_c=3.3$ fm for $^{11}$O from our DRHBc calculation, as it is not available in current DRHBc databases \cite{DRHBc22,DRHBc24}. We observe that the charge radii of $^{11}$Li/O are significantly off the experimental systematics, indicating a greater asymmetry than expected. Given that our prediction is based on the assumption of charge-density symmetry between mirror nuclei, the large deviation for $^{11}$O here is understandable. A similar argument applies for $^{11}$Be/N and $^{12}$Be/O mirror nuclei, whose $R_c$ values also deviate from the general systematics, as shown in Fig. \ref{fig:sym-rc}. The disagreements for $^{11}$N and $^{11,12}$O suggest that our method may also serve as a sensitive probe for detecting large charge-density asymmetry between mirror nuclei.
	
	\begin{figure}
	\includegraphics[angle=0,width=0.45\textwidth]{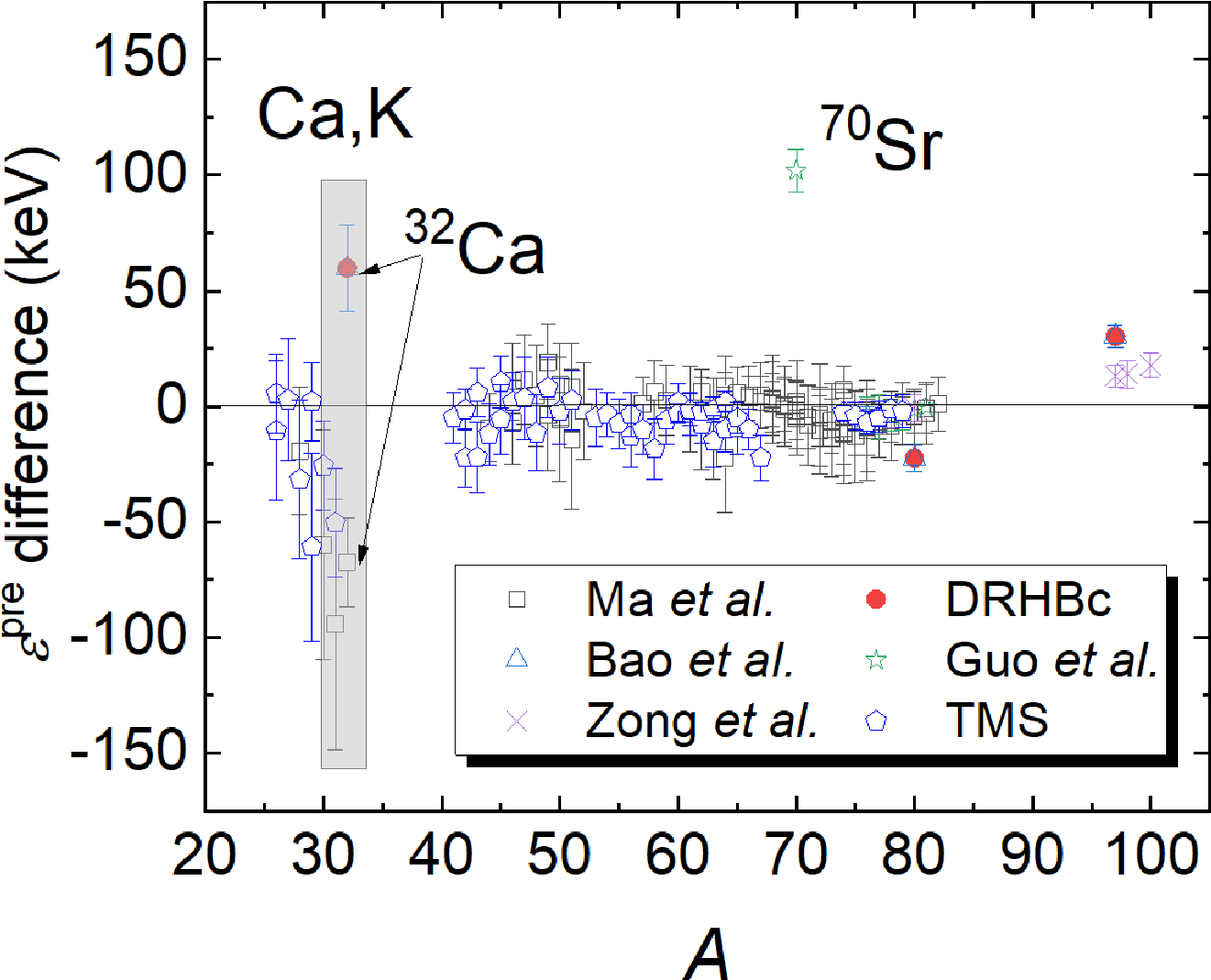}
	\caption{(Color online)
		Difference between the predicted binding energies per nucleon ($\varepsilon^{\rm pre}$) in this work and those from Ma $et~al.$ \cite{zong20}, the DRHBc databases \cite{DRHBc22,DRHBc24}, Bao $et~al.$ \cite{BAO2024}, Guo $et~al.$ \cite{FU2024} Zong $et~al.$ \cite{zong-self}, and the mass-surface (TMS) estimation in AME20 \cite{AME201}, These differences are calculated by subtracting our $\varepsilon^{\rm pre}$ values from the corresponding $\varepsilon^{\rm pre}$ values in other works. Our prediction uncertainty is determined as $\sqrt{\delta^2_\varepsilon+\sigma^2_\varepsilon}$ using Eq. (\ref{eq:ba-err}) and (\ref{eq:ba-unc}). The obvious inconsistency for some $A\sim 30$ K or Ca isotopes with $Z=19,20$ (especially $^{32}$Ca), and $^{70}$Sr is emphasized.
	}\label{fig:ba-pre}
	\end{figure}
	
	We predict binding energies per nucleon for some proton-rich nuclei that are not experimentally available, and compare them with those from other works \cite{zong20,DRHBc22,DRHBc24,BAO2024,FU2024,zong-self} in Fig. \ref{fig:ba-pre} by presenting the differences between their predictions and ours. Such differences are calculated by subtracting our $\varepsilon^{\rm pre}$ values from the corresponding $\varepsilon^{\rm pre}$ values in other works. Most of these differences in $\varepsilon^{\rm pre}$ are close to 0 within the $1\sigma$ confidence interval, indicating that our prediction generally agrees with others, except for some K and Ca isotope around $A\sim 30$, as well as $^{70}$Sr. For these exceptions, our predictions deviate from others' by $\sim 100$ keV. Notably, for $^{32}$Ca, Ma $et~al.$, Bao $et~al.$/DRHBc and our predictions show significant difference in mass estimates. We plot $R_c$s of $^{32}$Mg/Ca and $^{70}$Ge/Sr mirror pairs in Fig. \ref{fig:sym-rc}, where $R_c$s of $^{32}$Mg and $^{70}$Ge are from Rc13/21, and those of $^{32}$Ca, $^{70}$Sr are from DRHBc24 and our own DRHBc calculation, since their $R_c$s are experimentally unavailable. These $R_c$s of mirror nuclei deviate clearly from the regular systematics, showing much larger asymmetry than other mirror nuclei. Similar to the $^{11}$N and $^{11,12}$O cases in Fig. \ref{fig:ba20to20}, this large asymmetry affects our predictions for binding energies per nucleon of these exceptions. 
	
\begin{figure}
		\includegraphics[angle=0,width=0.45\textwidth]{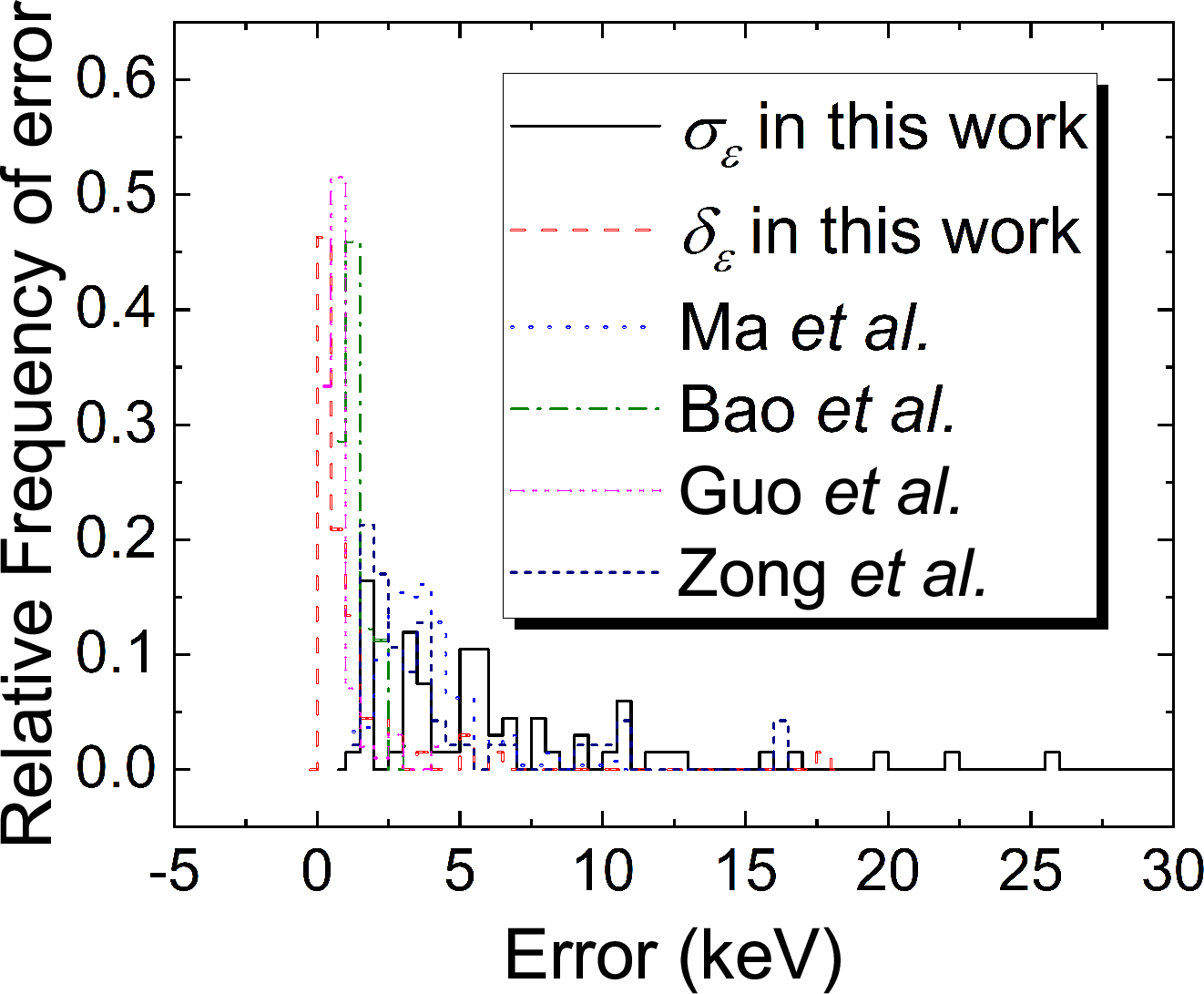}
		\caption{(Color online)
			Distribution of prediction uncertainty for experimentally unavailable $\varepsilon$ in this work, as well as in the predictions by in Ma $et~al.$ \cite{zong20}, Bao $et~al.$ \cite{BAO2024}, Guo $et~al.$ \cite{FU2024} and Zong $et~al.$ \cite{zong-self}. $\delta_\varepsilon$ is calculated using Eq. (\ref{eq:ba-err}), representing the uncertainty due to experimental input errors. $\sigma_\varepsilon$ is calculated using Eq. (\ref{eq:ba-unc}), corresponding to the modeling uncertainty in our $\eta$ estimation.
		}\label{fig:ba-unc-dis}
	\end{figure}

	For a comprehensive comparison between the mass-prediction uncertainties across different models, we present the uncertainty distributions of $\varepsilon^{\rm pre}$ from various local correlations \cite{zong20,BAO2024,FU2024,zong-self} and this work, for experimentally unavailable nuclei, in Fig. \ref{fig:ba-unc-dis}. In our perdition, the uncertainty due to experimental errors in inputted binding energy per nucleon and charge radius of corresponding mirror nucleus is typically around 1 keV, which is similar to the uncertainties found in other models. However, the modeling uncertainty from our $\tilde\eta$ estimation is widely spread, reaching up to $\sim$25 keV, which is the primary contributor to the uncertainty in our predictions. While the robustness of $\tilde\eta$ offers a stable prediction method, it also makes our correlation less sensitive to local structural detail, which can lead to significant uncertainty in the fitting parameters, thereby reducing the reliability of predictions made using this correlation.
	
	\subsection{RMS charge radii}
	
	Similarly, we can use the known binding energies per nucleon of a mirror pair to predict their RMS charge radii, by transforming Eq. (\ref{eq:eta-def}) into
	\begin{equation}\label{eq:rc-pre}
		R_c(Z,N)=\frac{\tilde\eta(Z,N)(N-Z)}{\varepsilon(Z,N)-\varepsilon(N,Z)}.
	\end{equation}
	The corresponding prediction uncertainty due to experimental input errors is given by
	\begin{equation}\label{eq:rc-err}
		\delta_{R_c(Z,N)}=\frac{
			R_c(Z,N)\sqrt{
				\delta^2_{\varepsilon(Z,N)}+\delta^2_{\varepsilon(N,Z)}
			}
		}
		{|\varepsilon(Z,N)-\varepsilon(N,Z)|},
	\end{equation}
	while the modeling uncertainty due to the imperfection of our $\eta$ estimation is
	\begin{equation}\label{eq:rc-unc}
		\sigma_{R_c(Z,N)}=\frac{R_c(Z,N)}{\tilde\eta(Z,N)}\sigma_{\tilde\eta(Z,N)}.
	\end{equation}
	The total uncertainty is then calculated as $\sqrt{\delta^2_{R_c}+\sigma^2_{R_c}}$. In this work, we predict the charge radii of 199 nuclei, which are listed along with their corresponding total uncertainties in the Supplemental Materials of this paper \cite{supp}.
	
\begin{figure}
		\includegraphics[angle=0,width=0.45\textwidth]{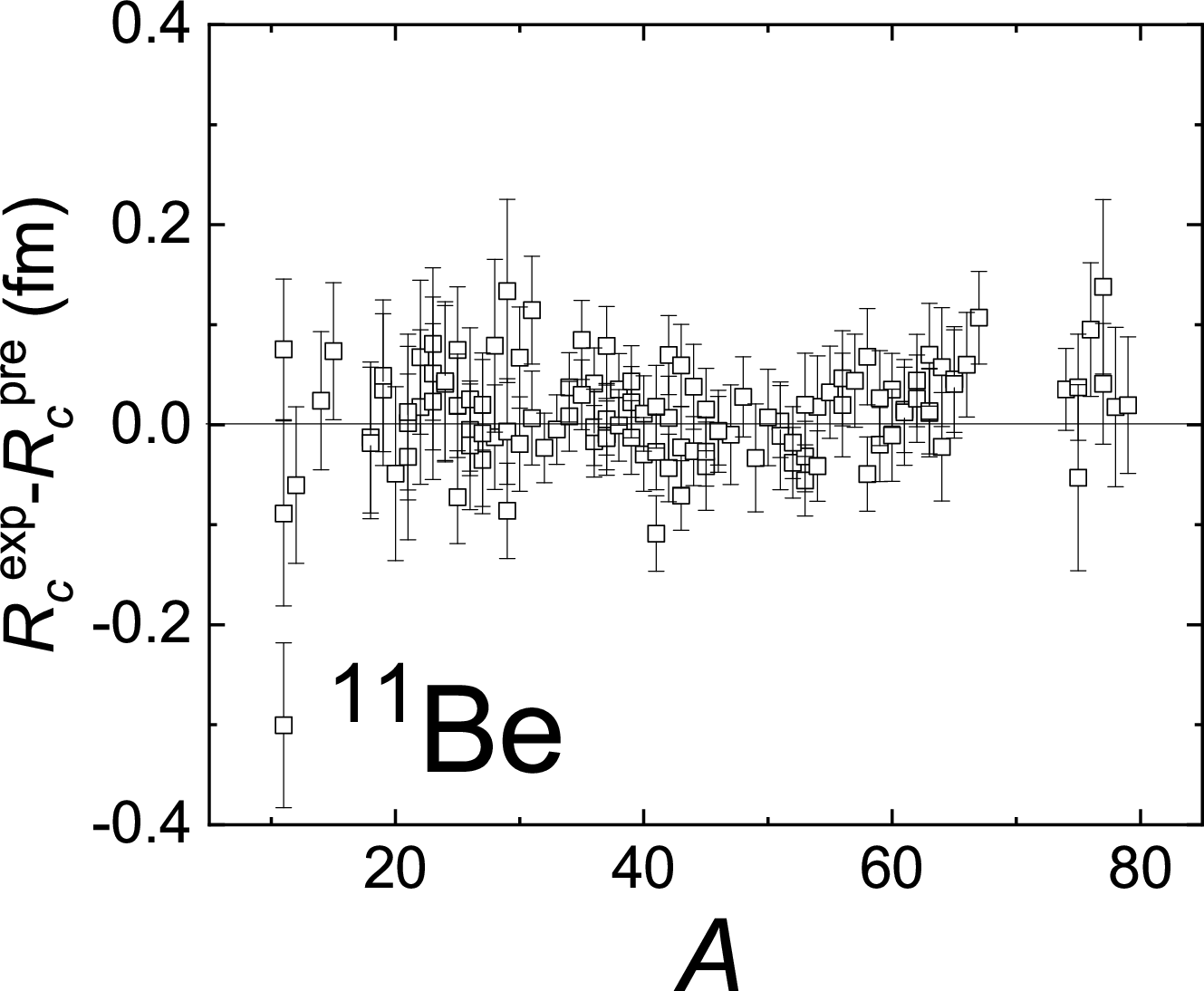}
		\caption{Difference between the ``predicted" charge radii using AME20 and Eq. (\ref{eq:rc-pre}), and experimental charge radii ($R^{\rm exp}_c$) from Rc13, Rc21, {\color{blue}and Refs. \cite{Sc-self,Sc-6,Sc-8,K-self}}, as a function of mass number $A$. The prediction error is calculated as $\sqrt{\delta^2_{R_c}+\sigma^2_{R_c}}$ using Eqs. (\ref{eq:rc-err}) and (\ref{eq:rc-unc}). The inconsistency for $^{11}$Be is highlighted.}\label{fig:rc-to20}
	\end{figure}

	Using Eq. (\ref{eq:rc-pre}), we can utilize the AME20 database to ``predict" some $R_c$ values available in Rc13/21, and compare them to experimental values from Rc13/21, by showing the difference between our ``predictions" and experimental charge radii in Fig. \ref{fig:rc-to20}. Our predictions mostly agree with experiments within the $1\sigma$ confidence interval, except for $^{11}$Be, which is the mirror nucleus of $^{11}$N, showing a clear inconsistency in the binding energy prediction in Fig. \ref{fig:ba20to20}. The RMSD for all available data is 0.054 fm, and excluding $^{11}$Be it is 0.047 fm, which is of the same order of RMSDs in Refs. \cite{DRHBc22,DRHBc24,bao76}, but is noticeably larger that that in Ref. \cite{bao75}. The accuracy of the predicted charge radii is not optimal, which is understandable. Our method considers the product of mass difference in mirror nuclei and corresponding charge radius as a robust observable, $\eta$. Therefore, the accuracy of both mass and charge radius predictions should be positive correlated. Since the accuracy of our mass prediction is not competitive, we don't expect our charge-radius predictions to be highly accurate either.
	
	Nevertheless, we still need to evaluate whether our predictions have a systematic deviation from experiments, or over-fitting problem, using the $\chi^2$ analysis. The reduced $\chi^2_{R_c}$ is defined as
	\begin{equation}
		\chi^2_{R_c}=\frac{1}{N-3}\sum\frac{(R^i_{c ~\rm exp}-M^i_{c ~\rm pre})^2}{\left(\sigma^i_{\rm exp}\right)^2+\left(\sigma^i_{\rm pre}\right)^2},
	\end{equation}
	where $R^i_{c \rm exp}$, $R^i_{c~\rm pre}$, $\sigma^i_{\rm exp}$ and $\sigma^i_{\rm pre}$ are the experimental and predicted charge radius of the $i$-th nucleus available, and corresponding experimental error and prediction uncertainty, calculated as $\sqrt{\delta^2_{R_c}+\sigma^2_{R_c}}$ using Eqs. (\ref{eq:rc-err}) and (\ref{eq:rc-unc}), respectively.
	For the entire dataset, $\chi^2_{R_c}=0.955$, and for those excluding $^{11}$Be, $\chi^2_{R_c}=0.850$, both of which are close to 1, suggesting that our prediction uncertainty is reasonably estimated, and that there is no systematic deviation in our predictions of charge radii.
	
	For $^{11}$Be, the predicted $R^{\rm pre}_c$ is larger than the experimental value by $\sim$0.3 fm. As discussed previously, the significant charge-density asymmetry in the $^{11}$Be/N mirror nuclei leads to a very different experimental $\eta_{\rm exp}$ compared to our estimation $\tilde\eta$, which explains this discrepancy for $^{11}$Be, as shown in Fig. \ref{fig:sym-rc}. Therefore, we believe that every failure in our prediction may correspond to an off-systematics behavior in the $R_c$ diagram of mirror nuclei as seen in  Fig. \ref{fig:sym-rc}.	
	
\begin{figure}
		\includegraphics[angle=0,width=0.45\textwidth]{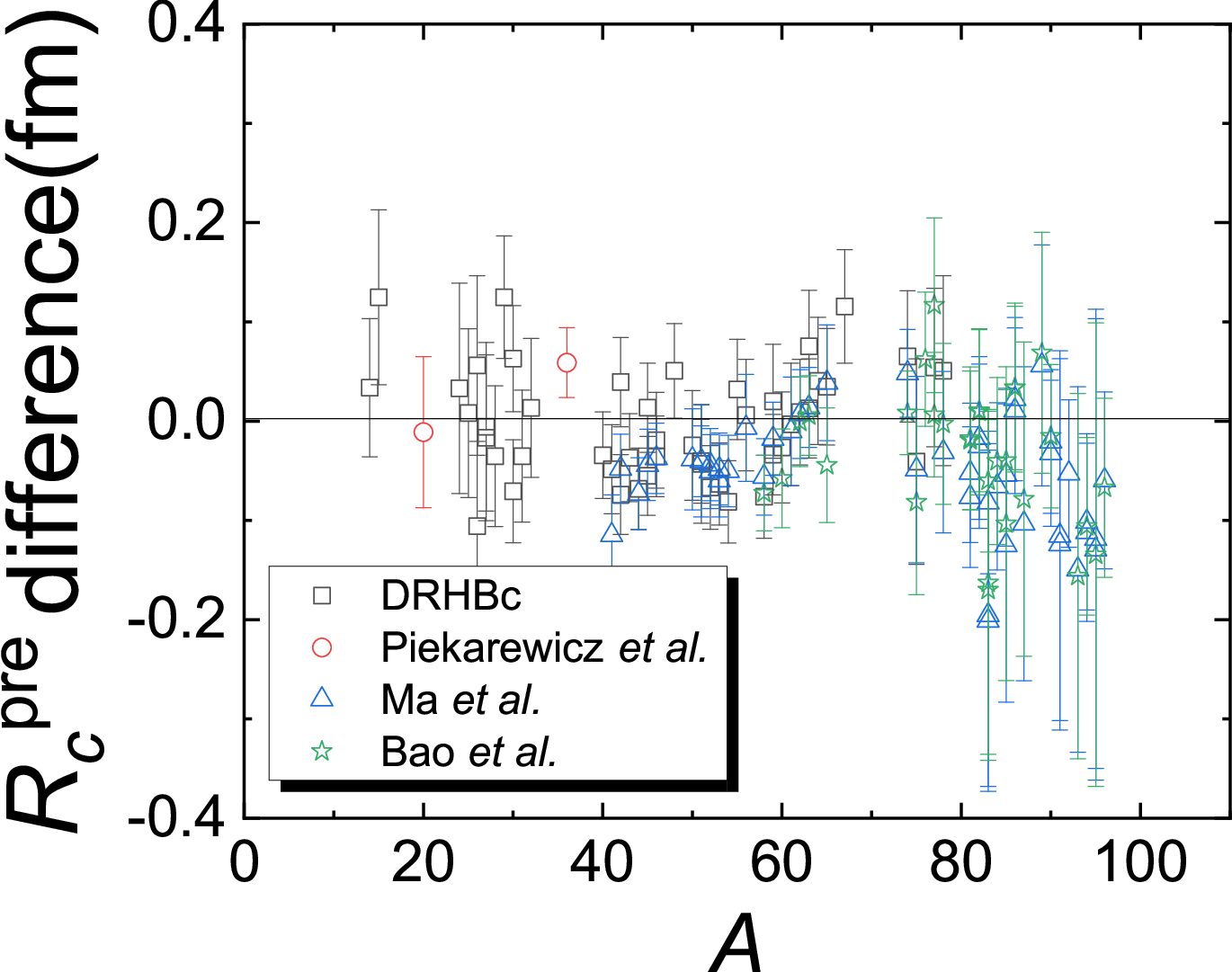}
		\caption{(Color online)
			Difference between $R^{\rm pre}_c$ from this work and  those from other works, including DRHBc databases \cite{DRHBc22,DRHBc24}, Piekarewicz $et~al.$'s \cite{bao71}, Ma $et~al.$'s \cite{bao76}, and Bao $et~al.$'s \cite{bao75} works, as a function of mass number $A$. The differences are calculated by subtracting our $R_c^{\rm pre}$ values from the corresponding $R_c^{\rm pre}$ values in other works.
		}\label{fig:rc-diff}
	\end{figure}

	We also use Eq. (\ref{eq:rc-pre}) to predict unknown $R_c$ ($R^{\rm pre}_c$) using AME20, and compare these predicted values with those from other works, by presenting the difference between $R^{\rm pre}_c$ from this work and those from literature, as shown in Fig. \ref{fig:rc-diff}. Such differences are calculated by subtracting our $R_c^{\rm pre}$ values from the corresponding $R_c^{\rm pre}$ values in other works. We observe reasonable agreement between our predictions and those from other works within $\sim$ 0.2 fm deviation, which is nearly the $1\sigma$ confidence interval.
	
	\subsection{Bound/unbound nature for inconsistencies}
	
	By analyzing all the inconsistencies between our predictions and experimental results, or other predictions, we identify a pattern: each inconsistency corresponds to a deviation from the regular systematics in the mirror $R_c$ plot, as shown in Fig. \ref{fig:sym-rc}. Each off-systematics data point involves a $Z>N$ nucleus with a significantly larger charge radius than its $Z<N$ mirror, which could feature a possible proton halo structure \footnote{See Refs. \cite{p-halo-b,p-halo-ne,p-halo-al-1,p-halo-al-2,p-halo-zhang} and references therein, for recent studies on the proton halo.}, if this $Z>N$ nucleus is weakly bound. Therefore, we are highly motivated to investigate every inconsistency from our predictions to determine whether the corresponding mirror nucleus is bound or not, i.e., whether its proton separation is positive or negative.
	
	\begin{table}
		\caption{Proton (or two-proton) separation energies (in MeV) for mirror nuclei that deviate from systematic trends in Fig. \ref{fig:sym-rc}. All data are based on the binding energies from AME20, except for $^{32}$/Mg/Ca and $^{70}$Ge/Sr, where the relevant binding energies are taken from DRHBc24 \cite{DRHBc24} and DRHBc calculations. Due to insufficient data, we present only the two-proton separation energy of $^{70}$Sr.}\label{tab:sp}
		\begin{tabular}{cc|ccccccccccccccccccccccccc}
			\hline\hline
			$Z<N$ & $S_p$ & $Z>N$ & $S_p$ & $S_{2p}$ \\
			$^{11}$Li & 15.76(9) & $^{11}$O & -1.6(4) &  \\
			$^{11}$Be & 20.16(1) & $^{11}$N & -1.378(5) &  \\
			$^{12}$Be & 22.939(2) & $^{12}$O & -0.36(1) &  \\
			$^{32}$Mg & 20.36(1) & $^{32}$Ca & 3(1) &  \\
			$^{70}$Ge & 8.523(1) & $^{70}$Sr &  & -1.5(5) \\
			\hline\hline
		\end{tabular}
	\end{table}
	
	We list the proton (or two-proton) separation energies of these off-systematics mirrors in Table \ref{tab:sp}. These off-systematics mirrors mostly exhibit an unbound nature in their $Z>N$ nuclei, with negative proton separation energy beyond the proton drip line, except for $^{32}$Mg/Ca mirror. The bound ground state of $^{32}$Ca can be attributed to the $Z=20$ magic number, which is unlikely to produce a proton halo. Thus, the robust correlation proposed so far does not identify a candidate with a proton halo, but we believe it has the potential to do so, as more binding energies and charge radii of proton rich nuclei become available from future experiments. 
	
	Beyond the asymmetry in the $^{32}$Mg/Ca mirror pair, the robustness of the $Z=20$ magic number emerges becomes evident. The off-systematics behavior of charge radii in this mirror pair indicates that the charge density of $^{32}$Ca, associated with the $Z=20$ magic number, is markedly different from that that of $^{32}$Mg, where the proton Fermi surface is at $Z=12$ in the middle of the $sd$ shell. This difference results in an inaccurate prediction of the binding energy for $^{32}$Ca, as previously demonstrated. Furthermore, a large charge radius may correspond to a less bound structure, as observed in other bound-unbound mirror nuclei listed in Table \ref{tab:sp}, where the $Z>N$ nuclei consistently exhibit larger charge radii and negative proton separation energies according to Fig. \ref{fig:sym-rc} and Table \ref{tab:sp}. Therefore, the relatively larger charge radius of $^{32}$Ca may suggest an unbound ground state. However, $^{32}$Ca still has bound ground state and a positive proton separation energy, highlighting the robustness of $Z=20$ magic number even when the neutron number is reduced to $N=12$, and the charge distribution is significantly spread.
	
	For other bound-unbound mirror nuclei in Table \ref{tab:sp}, we have shown that unbound protons can potentially enlarge the charge radius, increase the asymmetry in the charge density, and reduce the reliability of our predictions. Therefore, to accurately predict the masses of the unbound nuclei in a mirror pair, different proton densities may need to be introduced for the bound and potentially unbound nuclei of a mirror pair, as was done in Ref. \cite{zong-self}. Specifically, we should consider using different $R$ values and $a/R$ ratios for both $Z>N$ and $Z<N$ nuclei of a mirror pair. However, we have not yet found an experimental survey of the $a$ and $R$ parameters for $Z>N$ nuclei, making it ungrounded to introduce another set of $R$ and $a/R$ values at this time. 
	
	{\color{blue}
	\subsection{$A$ dependence of charge radii}
	
	It's known that exotic nuclei, especially those near the proton dripline, do not follow the usual regularity of $A$ dependence as stable nuclei do. It is interesting to examine the $A$ dependence of our predicted charge radii, and to analyze how these predicted charge radii uniquely evolve with mass number near the proton dripline.
	
	\begin{figure}
		\includegraphics[angle=0,width=0.45\textwidth]{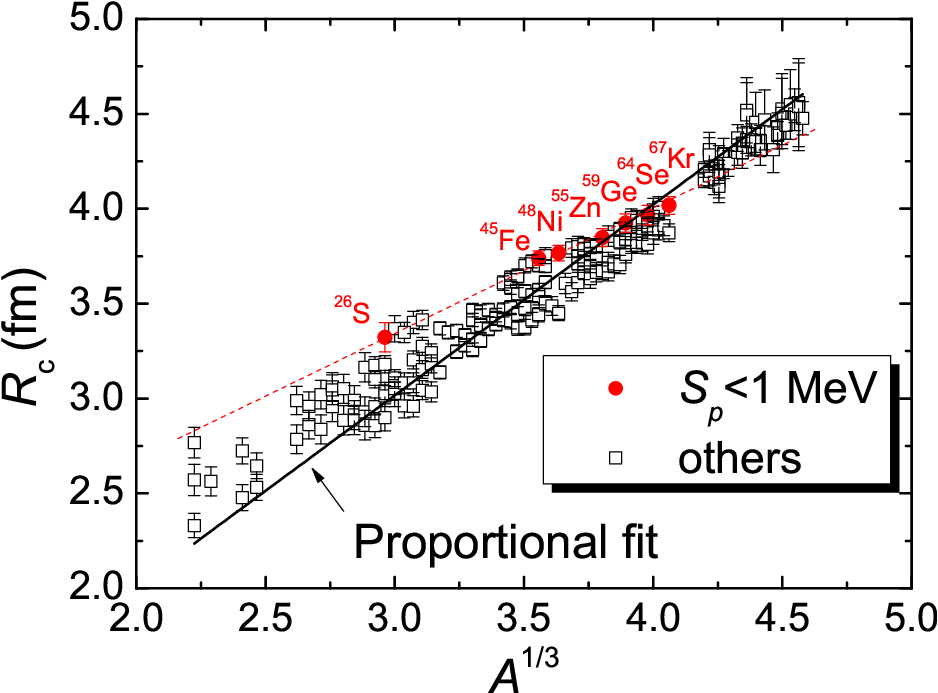}
		\caption{(Color online) \color{blue} 			Predicted charge radii as a function of $A^{1/3}$. A proportional fit is applied to all predicted charge radii, and represented by the solid black line, illustrating the conventional rule of $R_c\propto A^{1/3}$ regularity. Nuclei near the proton dripline, with an experimental proton separation energy of $S_p<1$ MeV \cite{AME202}, are highlighted with red dots. A dashed red line is included to guide the eye, and to  schematically illustrate the mass dependence of charge radii near the proton dripline. 
		}\label{fig:rc-a13}
	\end{figure}
	 
	We plot our predicted charge radii against $A^{1/3}$ in Fig. \ref{fig:rc-a13}. A proportional fit is applied, demonstrating the general regularity of $R_c\propto A^{1/3}$ for the predicted charge radii, which is expected due to the incompressibility of nuclear matter. On the other hand, we highlight the nuclei near the proton dripline, specifically those with experimental proton separation energies \cite{AME202} of less than 1 MeV, in Fig. \ref{fig:rc-a13}. We find that the predicted radii near the proton dripline indeed follow a different systematics compared to the global $R_c\propto A^{1/3}$ regularity. Therefore, we believe our predictions reasonably establish the known and obvious difference in $A$ dependence between nuclei near proton dripline and those farther away, further validating our predictions.
	
	\begin{figure}
		\includegraphics[angle=0,width=0.45\textwidth]{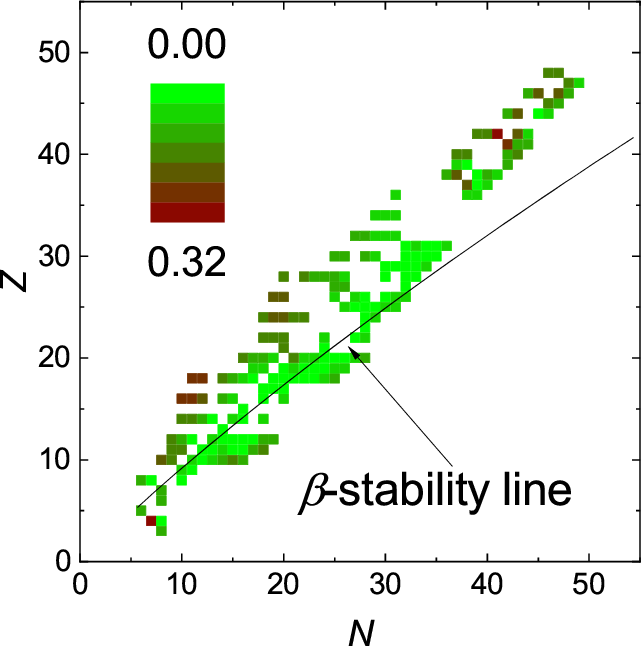}
		\caption{(Color online) \color{blue}
			Deviation of our predicted charge radii from Eq. (\ref{eq:rc-stable}), which is considered to be the empirical formula of charge radii along the line of $\beta$ stability. 
		}\label{fig:land-rc}
	\end{figure}
	
	Moreover, the charge radii along the line of $\beta$ stability follows
	\begin{equation}\label{eq:rc-stable}
		R_c=\left(r_0+\frac{r_1}{A^{2 / 3}}+\frac{r_2}{A^{4 / 3}}\right) \times A^{1 / 3}, 
	\end{equation} 
	where $r_0=0.707$, $r_1=1.11$ and $r_2=-0.55$ are obtained from the least-squares fitting of Eq. (8) in Ref. \cite{rc13}. In the nuclide landscape of Fig. \ref{fig:land-rc}, we present the deviation of our predicted charge radii from Eq. (\ref{eq:rc-stable}) using color mapping, which provides another perspective on the difference in the $A$ dependence of charge radii between proton dripline nuclei and stable nuclei in our predictions. It can be seen that along the $\beta$-stability line, the deviation is quit small, whereas as approaching the proton dripline, the deviations become increasingly significant. This also demonstrates that the charge radii of proton-dripline nuclei exhibit a different $A$ dependence compared to those of stable ones, as expected.
	}
			
	\section{summary}\label{sec:sum}
	 {\color{blue} In this work, we propose a new, robust and nontrivial correlation between binding energies and charge radii of mirror nuclei.}
	 This correlation is governed by an observable $\eta$ defined in Eq. (\ref{eq:eta-def}), which can be estimated using Eq. (\ref{eq:z-dep}), independently of experiment data, relying only on the mass number $A$ and proton number $Z$. 
	
	The numerical analysis suggests that $\eta$ varies little with the fluctuations in the nuclear geometry radius and charge surface thickness. It also remains stable across different mass regions, as confirmed by experimental survey. Therefore, it provides a robust correlation that can be applied to predict the binding energies and charge radii of proton rich nuclei. 
	
	{\color{blue}
	We note that the proposed correlation is derived analytically using the liquid drop model and the Fermi model for nuclear charge density. Both models are global frameworks that utilize empirical formulas based on the experimental data of the near-stable nuclei. However, this correlation also apply across a wide range of nuclide landscape, especially for proton-rich nuclei, illustrating the robustness of this correlation.
	}
	
	{\color{blue}
	We also emphasize that the correlation proposed here is not trivial. Although the binding energy and charge radius are both related to mass number and charge number, which might suggest an apparent natural correlation, the reality is more complex. These two observables not only depend on mass number and charge number, but also are significantly influenced by the shape of charge density, exchange symmetry of identical protons and spin-orbit coupling within the Fermi model and the liquid drop model. Therefore, establishing an analytical correlation between them necessitates sophisticated and strong interplay between these two models, including the formalism consistency implied in Sec. \ref{sec:eta}, the specific parameter regularity illustrated in Fig. \ref{fig:aR_A} and the global invariance of $\eta$ highlighted in Fig. \ref{fig:eta-A-exp}. Hence, we do not take this correlation for granted.
	}
	
	Using this correlation, we predict the binding energies per nucleon of 199 nuclei and the charge radii of 197 nuclei, which are listed along with their corresponding uncertainties in the Supplement Material of this paper \cite{supp}. The predicted results agree with current experimental data and other theoretical predictions, with few exceptions. The inclusion of unbound nuclei may account for the relatively large RMSD in the mass and charge-radius predictions, making them less competitive in terms of accuracy. However, the reduced $\chi^2$ analysis suggests that our uncertainty evaluation reasonably consists with and justifies such large RMSD. The performance of our prediction method could be improved by introducing different $R$ values and $a/R$ ratios for $Z>N$ and $Z<N$ nuclei of mirror pairs. However, it is currently challenging to incorporate an additional set of Fermi parameters for $Z>N$ nuclei, due to the lack of experimental foundation.
	
	We focus on all the inconsistencies between our predictions and experimental results, or other predictions, which are associated with the usually large asymmetry in the charge densities of mirror nuclei. Such pronounced asymmetry may indicate exotic structures of $Z>N$ nuclei in a mirror pair, such as the magic nature observed in the $^{32}$Mg/Ca mirror pair, or the potential proton halo structure. Therefore, we believe that the proposed correlation could serve as a sensitive probe for detecting the asymmetry or difference in the charge structures of mirror nuclei.
	
	{\color{blue}
	We also examine the mass dependence of predicted charge radii, and highlight significant difference in mass dependence between nuclei near the proton dripline and those farther away. This systematic difference is consistent with previous understandings of the charge radii of exotic nuclei, which further supports our prediction method.
	}
	
	\begin{acknowledgments}
		We are grateful to Dr. Y. M. Zhao for inspiring us regarding the proton halo structure, and to Mr. S. T. Guo for discussions on modeling uncertainty. This work is supported by the National Natural Science Foundation of China (Grants No. 12105234 and No. 12305125), and the Sichuan Science and Technology Program (Grant No. 2024NSFSC1356).
	\end{acknowledgments}
	
	%
	
\end{document}